\documentstyle[aps,epsf]{revtex}
\begin{document}
\title{Precise nondivergent analytic formulas for the radiative corrections to the $%
\beta $ energy spectrum in hyperon semileptonic decays over the entire
Dalitz plot}
\author{S. R. Ju\'{a}rez W. and F. Guzm\'{a}n A.}
\address{Escuela Superior de F\'{\i }sica y Matem\'{a}ticas. Instituto\\
Polit\'{e}cnico Nacional. Edificio 9, Unidad Prof. Adolfo L\'{o}pez Mateos, 
\\
Col. Lindavista, C.P. 07738 M\'{e}xico D.F., Mexico}
\maketitle

\begin{abstract}
Very accurate analytical expressions for the radiative corrections of
unpolarized hyperons semileptonic decays of charged and neutral baryons have
been obtained in the recent past. Some of these formulas contain logarithmic
singularities at the edges of the Dalitz plot for the three and four-body
decays. These singularities are analyzed and integrated analytically to
obtain new divergentless formulas for the energy spectrum of the produced $%
\beta $ particle. The new equations contain terms of the order $\alpha $
times the momentum transfer, are applicable to any beta decay process and
are suitable for a model-independent experimental analysis.

13.40.Ks, 13.30.Ce
\end{abstract}

%\date{\today}

\section{INTRODUCTION}

The subject of hyperon decays is of unquestionable importance in high energy
physics. The precise description of the semileptonic weak decays and in
particular the neutron decay, is relevant in cosmology, astrophysics, solar
physics, the solar neutrino problem, and in other areas of particle physics.
A precise formula for the energy spectrum of the decay products (fermions)
in the hyperon semileptonic decay (HSD) requires the knowledge of the {\it %
radiative corrections} (RC) in the whole region of the Dalitz plot (DP). The
total decay rate can be computed directly through analytical formulas for
the RC, if these formulas do not contain any divergences. The subject of
this paper is to show how the logarithmic kinematical divergences, that
appeared in the previously calculated model-independent bidimensional
distributions in Refs.~\cite{95}-\cite{96n}, obtained in an analytical way
for the RC to baryon $\beta $ decays in the three body region (TBR) and the
four body region (FBR) of the DP, are canceled after performing the
integration over the energy of the final hyperon that emerges in the
process. The new results are valid in any charged or neutral hyperon beta
decay. The important feature of the obtained analytical forms of the RC's is
that they are written in a simple form as products of two factors. One of
these factors is a model-independent function and the other one does depend
on the form factors whose determination (through the experimental data) is
useful to obtain information about the underlying interactions in the decay
processes, basic symmetries and the internal structure of hadrons. The new
formulas are suitable for a direct evaluation of the RC's for any event at
any point of the allowed physical region.

The structure of this paper is as follows. In Sec.~II we exhibit the
kinematical region in which the three and the four body decays take place.
All the amplitudes, the one without any RC, the bremsstrahlung and the
virtual ones, for the charged and neutral HSD, with all the $\alpha q/\pi
M_{1}$ terms included, are displayed in this section. The bidimensional
observable energy distributions in the FBR and TBR are described for the
charged and neutral HSD processes in Sec.~III . We devote Sec.~IV to display
the formulas for the energy spectrum of the emitted charged lepton in a
non-divergent form for the pure bremsstrahlung correction in the FBR, and
the formulas for the energy spectrum in the TBR with both kinds of RC 's
(the combination of the bremsstrahlung and the virtual contributions). In
Sec.~V we present our final formulas in each region for the charged and the
neutral decay processes. Conclusions are given in Sec.~VI, with the formulas
that contain all the up-to-date improvements and new features for the energy
spectrum, together with a comparison with other previously published
numerical values assuming their values for the form factors. Several graphs
for the partial radiative corrections (the Coulomb effect is included for
the neutral HSD process at the TBR) and for the complete electron spectrum
are presented in order to illustrate our results. Finally, for
self-containment, we include several appendices with the definitions of the
coefficients and model-independent functions that appear in the final
analytical results, among other relevant relations. The procedure of
integration with the explicit cancellations of the kinematical divergences
is also included.

\section{KINEMATICS AND AMPLITUDES}

We start with the presentation of the main features and the notation to
describe the decay processes we are interested in: 
\begin{equation}
A^{s}(p_{1})\rightarrow B(p_{2})+e^{-}(\ell )+\bar{\nu}_{e}(p_{\nu }),
\label{uno}
\end{equation}
where the emission of a virtual or a real photon $\gamma (k)$ takes place.
The emission of the photon is described as a radiative correction to the
semileptonic decay of the hyperon. $A^{s}$ corresponds to the neutral $(s=n)$
or charged $(s=c)$ decaying baryon, $B$ is the produced baryon, $e^{-}$ and $%
\bar{\nu}_{e}$ denote the lepton and its antineutrino counterpart,
respectively. The four-momenta and masses of the particles involved in the
baryon semileptonic decay are denoted by 
\[
p_{1}=(E_{1},\vec{p}_{1}),\quad p_{2}=(E_{2},\vec{p}_{2}),\;\;\ell =(E,\vec{%
\ell}),\quad p_{\nu }=(E_{\nu },\vec{p}_{\nu }),\;\;\mbox{and}\;k=(k_{0},%
\vec{k}), 
\]
and by $M_{1},M_{2},m,m_{\nu }$, and $m_{k}$ , respectively. We assume
throughout this paper that $m_{\nu }=0$, and $m_{k}=0$ for the real photons.

The conservation of energy and momentum determines the physical region where
the HSD takes place. For the three body decay as in the HSD (without any
radiative correction) the physical region is bounded by the hyperon minimal
and a maximal energy $\left( E_{2}^{\min },\,E_{2}^{\max }\right) $ for each
value of $E$. When a real photon is considered as an additional product of
the decay the physical four body region for this process contains the former
region (TBR) and an additional portion (FBR) with energies below the $%
E_{2}^{min}(E)$ in such a way that $M_{2}<E_{2}<E_{2}^{min}$ for $m<E<E_{c}$
where $E_{c}$ is the maximum electron energy when the final $B$-hyperon is
at rest. The upper boundary of $E_{2}$ , of the region (FBR) where the three
body decay is forbidden, is evaluated with the condition $k=0,$ this means,
without the emission of a real photon. Physically this corresponds to the
case in which the neutrino balances the total momentum of the residual
hyperon and electron that are emitted in colinear directions.

Consequently, the Dalitz Region for the process in Eq.~(\ref{uno}), in the
rest frame of $A^{s}$, is defined at the FBR by 
\begin{equation}
M_{2}\leq E_{2}\leq E_{2}^{\min },\,\,\,\,\,\,m\leq E\leq E_{c},\quad
\,\,\,\,\,\,\,\,E_{c}=\frac{\left( M_{1}-M_{2}\right) ^{2}+m^{2}}{2\left(
M_{1}-M_{2}\right) },  \label{klim1}
\end{equation}
and at the TBR by 
\begin{equation}
E_{2}^{\min }\leq E_{2}\leq E_{2}^{\max },\,\,\,\,\,\,m\leq E\leq
E_{m},\,\,\,\,\,\,\,\,\,E_{m}=\frac{M_{1}^{2}-M_{2}^{2}+m^{2}}{2M_{1}},
\label{klim2}
\end{equation}
where 
\begin{equation}
E_{2}^{\stackrel{\scriptstyle\max }{\scriptstyle\min }}=\frac{1}{2}\left(
M_{1}-E\pm {\beta E}\right) +\frac{M_{2}^{2}}{2\left( M_{1}-E\pm {\beta E}%
\right) },\,\,\,\,\mbox{and}\,\,\,\,\left| \vec{\ell}\right| ={\beta E.}
\label{klim0}
\end{equation}
The $z$-axis is chosen along the electron three-momentum and the $x$-axis
oriented so that the final baryon three-momentum is in the first or fourth
quadrants of the $x-z$ plane. To be more explicit, 
\begin{equation}
p_{1}=(M_{1},0),\,\,\,\,\,\,\,\,\,\,\,\,\,\,\,\,\,\,\,\,\,\,\,\,\vec{\ell}={%
\beta E}\,(0,\,0,\,1),\qquad \vec{p}_{2}=\left| \vec{p}_{2}\right| \;\left( 
\sqrt{\left( 1-y^{2}\right) },\;\,0,\;\;y\;\right)
.\,\,\,\,\,\,\,\,\,\,\,\,\,\,
\end{equation}
In the TBR, the radiative correction is due to a virtual or a real photon
emission, meanwhile in the FBR the emitted photon is a real photon
exclusively, i.e., the radiative correction is due only to the
brems\-strah\-lung effect in the semileptonic decay.

The uncorrected matrix element $M_{0}$ (without any emission of a photon at
the TBR) for the decay in Eq.~(\ref{uno}) is given by the product of the
matrix elements of the baryonic weak current and of the leptonic current: 
\begin{equation}
M_{0}=\frac{G_{v}}{\sqrt{2}}\bar{u}_{B}W_{\mu }u_{A}\bar{u}_{\ell }O_{\mu }%
{\it v}_{\nu }\quad ,
\end{equation}
where $G_{v}=G_{\mu }V_{ij}$ and $G_{\mu }$ is the muon decay coupling
constant, $V_{ij}$ is the corresponding Cabibbo-Kobayashi-Maskawa matrix
element, and 
\begin{equation}
W_{\mu }=f_{1}(q^{2})\gamma _{\mu }+\frac{f_{2}(q^{2})}{M_{1}}\sigma _{\mu
\nu }q_{\nu }+\frac{f_{3}(q^{2})}{M_{1}}q_{\mu }+\left[ g_{1}(q^{2})\gamma
_{\mu }+\frac{g_{2}(q^{2})}{M_{1}}\sigma _{\mu \nu }q_{\nu }+\frac{%
g_{3}(q^{2})}{M_{1}}q_{\mu }\right] \gamma _{5},\,\,\,%
\mbox{and \thinspace \thinspace \thinspace
\thinspace }O_{\mu }=\gamma _{\mu }(1+\gamma _{5}).
\end{equation}
The $q=p_{1}-p_{2}$ denotes the four-momentum transfer.

To obtain the precise decay rates, one has to consider all the involved
amplitudes of the processes with and without RC. It has been shown in Ref.~%
\cite{89M} that the order~$\alpha $ amplitude can be obtained in a model
independent fashion by using the Low-theorem, Refs.~\cite{Low}, \cite{Chew}.

For completeness, we reproduce here the model independent brems\-strah\-lung
amplitudes, in terms of the Dirac form factors, given for the charged HSD in
Eqs.~(18)-(20) of Ref.~\cite{91}, and for the neutral HSD in Eq.~(23) of
Ref.~\cite{93}.

The total {\it brems\-strah\-lung} transition amplitudes for the charged HSD 
$\,(s=c)$ and the neutral HSD $(s=n)$ decays in the entire Dalitz plot
region are

\begin{equation}
M_{B}^{s}=M_{1}^{s}+M_{2}^{s}+M_{3}^{s},\quad \quad \mbox{\quad
with\quad }
\end{equation}
\[
M_{1}^{c}=eM_{0}\left( \frac{\epsilon \cdot \ell }{\ell \cdot k}\;-\frac{%
\epsilon \cdot p_{1}}{p_{1}\cdot k}\right)
,\,\,\,\,\,\,\,\,\,\,\,\,\,\,M_{2}^{c}=\frac{eG_{v}}{\sqrt{2}}\epsilon _{\mu
}\bar{u}_{B}W_{\lambda }u_{A}\bar{u}_{\ell }\frac{\gamma _{\mu }k\!\!\!/}{%
2\ell \cdot k}O_{\lambda }v_{\nu } 
\]
\begin{eqnarray}
M_{3}^{c} &=&\frac{G_{v}}{\sqrt{2}}\bar{u}_{\ell }O_{\lambda }v_{\nu
}\epsilon _{\mu }\bar{u}_{B}\left[ \frac{eW_{\lambda }k\!\!\!/\gamma _{\mu }%
}{2p_{1}\cdot k}\right. -\kappa _{1}W_{\lambda }\frac{p_{1}\!\!\!\!\!/+M_{1}%
}{2p_{1}\cdot k}\sigma _{\mu \nu }k_{\nu }+\kappa _{2}\sigma _{\mu \nu
}k_{\nu }\frac{p_{2}\!\!\!\!\!/+M_{2}}{2p_{2}\cdot k}W_{\lambda }  \nonumber
\\
&&\ +e\left( \frac{p_{1\mu }k_{\lambda }}{p_{1}\cdot k}-g_{\mu \lambda
}\right) \,\left( \frac{f_{3}-f_{2}}{M_{1}}+\gamma _{5}\frac{g_{3}-g_{2}}{%
M_{1}}\right) +e\left( \frac{p_{1\mu }k_{\nu }}{p_{1}\cdot k}-g_{\mu \nu
}\right) \,\left. \left( \sigma _{\lambda \nu }+g_{\lambda \nu }\right)
\left( \frac{f_{2}+g_{2}\gamma _{5}}{M_{1}}\right) \right] u_{A},
\end{eqnarray}
and 
\[
M_{1}^{n}=eM_{0}\left( \frac{\epsilon \cdot \ell }{\ell \cdot k}\;-\frac{%
\epsilon \cdot p_{2}}{p_{2}\cdot k}\right) ,\qquad M_{2}^{n}=\,\,M_{2}^{c}, 
\]
\begin{eqnarray}
M_{3}^{n} &=&\frac{G_{v}}{\sqrt{2}}\bar{u}_{\ell }O_{\lambda }v_{\nu
}\epsilon _{\mu }\bar{u}_{B}\left\{ -\frac{e\gamma _{\mu }k\!\!\!/W_{\lambda
}}{2p_{2}\cdot k}\right. -\kappa _{1}W_{\lambda }\frac{p_{1}\!\!\!\!\!/+M_{1}%
}{2p_{1}\cdot k}\sigma _{\mu \nu }k_{\nu }+\kappa _{2}\sigma _{\mu \nu
}k_{\nu }\frac{p_{2}\!\!\!\!\!/+M_{2}}{2p_{2}\cdot k}W_{\lambda }\, 
\nonumber \\
&&+e\left( \frac{p_{2\mu }k_{\rho }}{p_{2}\cdot k}-g_{\mu \rho }\right)
\left. \left[ \left( \frac{f_{2}+g_{2}\gamma _{5}}{M_{1}}\right) \sigma
_{\lambda \rho }+g_{\lambda \rho }\left( \frac{f_{3}+g_{3}\gamma _{5}}{M_{1}}%
\right) \right] \right\} u_{A}.
\end{eqnarray}
$\kappa _{1}$ and $\kappa _{2}$ are the anomalous magnetic moments of A$^{s}$
and B, respectively. $\epsilon _{\mu }$ is the photon polarization
four-vector.

The total {\it virtual} transition amplitudes (present only at the TBR{\bf )}
with terms up to order of $\alpha q/\pi M_{1}$ can be written as 
\begin{equation}
M_{V}^{s}=M_{0}^{\prime }+M_{\upsilon }^{s},
\end{equation}
\noindent where $M_{0}^{\prime }$ is the uncorrected matrix element in terms
of effective form factors Ref.~\cite{80}.

Explicitly, for the charged HSD case 
\begin{equation}
M_{\upsilon }^{c}=\frac{\alpha }{2\pi }\left[ M_{0}\Phi _{c}+M_{p_{1}}\Phi
_{c}^{\prime }\right] ,
\end{equation}
where 
\begin{eqnarray}
\Phi _{c} &=&2\left[ \frac{1}{\beta }{\mbox{arctanh}}\left( {\beta }\right)
-1\right] \ln \left| \frac{\lambda }{m}\right| -\frac{1}{\beta }\left( {%
\mbox{arctanh}}\left( {\beta }\right) \right) ^{2}+\frac{1}{\beta }L\left[ 
\frac{2\beta }{1+\beta }\right] -\frac{1}{\beta }L\left[ \frac{2\beta }{%
M_{1}/E-1+\beta }\right] +\frac{3}{2}\ln \left| \frac{M_{1}}{m}\right| 
\nonumber \\
&&+\frac{1}{\beta }{\mbox{arctanh}}\left( {\beta }\right) \left[ 1+\frac{%
E\left( 1-\beta ^{2}\right) }{M_{1}-2E}\right] -\frac{1}{\beta }\ln \left[ 1-%
\frac{2\beta }{M_{1}/E-1+\beta }\right] \left[ \ln \left| \frac{M_{1}}{m}%
\right| -{\mbox{arctanh}}\left( {\beta }\right) \right] -\frac{11}{8},
\label{pt2}
\end{eqnarray}
\begin{equation}
\Phi _{c}^{\prime }=\frac{1-\beta ^{2}}{\beta }\left[ -{\mbox{arctanh}}%
\left( {\beta }\right) \left[ 1+\frac{E}{M_{1}-2E}\right] +\frac{\beta E}{%
M_{1}-2E}\ln \left| \frac{M_{1}}{m}\right| \right] ,  \label{fcpp}
\end{equation}
with 
\begin{equation}
L(x)=\int_{0}^{x}\frac{\ln \left( 1-t\right) }{t}dt\,\,\,\,{%
\mbox{is the
Spence function},\,\,\,\,\,\,\,\,\,\,\,\mbox{and}\,\,\,\,\,\,\,\,\,\,}%
M_{p_{1}}=\frac{G_{v}}{\sqrt{2}}\frac{E}{mM_{1}}\overline{u}_{B}W_{\rho
}u_{A}\overline{u}_{l}\,p_{1}\!\!\!\!\!/\,\,O_{\rho }v_{\nu }.
\end{equation}
The formulas for the neutral HSD are: 
\begin{equation}
M_{\upsilon }^{n}=\frac{\alpha }{2\pi }[M_{0}\Phi _{n}+M_{p_{2}}\Phi
_{n}^{\prime }],\,\,\,{\mbox{where}\,\,\,}M_{p_{2}}=\frac{G_{V}}{\sqrt{2}}%
\frac{l\cdot p_{2}}{m(l+p_{2})^{2}}\overline{u}_{B}W_{\rho }u_{A}\overline{u}%
_{l}p_{2}\!\!\!\!\!/\,\,O_{\rho }v_{\nu }.
\end{equation}
The $\Phi _{n}$ and $\Phi _{n}^{\prime }$ are given in Ref.~\cite{93},
Eqs.~(8)-(10), and can be written as

\begin{equation}
\Phi _{n}=\Phi _{n}^{IR}(E,E_{2},\lambda )+\Phi _{n}^{ND}(E,E_{2})+\Phi
_{Coulomb}(E,E_{2})  \label{virn}
\end{equation}
where 
\begin{equation}
\Phi _{n}^{IR}(E,E_{2},\lambda )=2\left[ \frac{1}{\beta _{N}}{\mbox{arctanh}}%
\!\left( \beta _{N}\right) -1\right] \ln \left| \frac{\lambda }{m}\right|
,\,\,\,\,\,\,\,\,\,\,\,\,\,\,\,\,\Phi _{Coulomb}(E,E_{2})=\frac{\pi ^{2}}{%
\beta _{N}},  \label{ficou}
\end{equation}
\begin{eqnarray}
\Phi _{n}^{ND}(E,E_{2}) &=&-\frac{1}{\beta _{N}}\left[ {\mbox{arctanh}}%
\!\left( \beta _{N}\right) \right] ^{2}+\frac{1}{\beta _{N}}L\left[ \frac{%
\Delta _{V}}{x_{2}^{+}}\right] -\frac{1}{\beta _{N}}L\left[ \frac{-\Delta
_{V}}{1-x_{2}^{+}}\right] +\frac{1}{\beta _{N}}{\mbox{arctanh}}\!\left(
\beta _{N}\right) \left[ \frac{M_{2}^{2}+\left( 1+\beta _{N}^{2}\right) a}{%
2\left( l+p_{2}\right) ^{2}}\right]  \nonumber \\
&&+\frac{3}{2}\ln \left[ \frac{M_{2}}{m}\right] -\frac{11}{8}-\frac{m^{2}}{%
\left( l+p_{2}\right) ^{2}}\ln \left[ \frac{M_{2}}{m}\right] -\frac{1}{\beta
_{N}}\ln \left[ 1+\frac{\Delta _{V}}{1-x_{2}^{+}}\right] \left[ \ln \left[ 
\frac{M_{2}}{m}\right] -{\mbox{arctanh}}\!\left( \beta _{N}\right) \right] ,
\end{eqnarray}
\begin{equation}
\Phi _{n}^{\prime }=\frac{1}{\beta _{N}}\left( 1-\beta _{N}^{2}\right)
\left[ -\left( 1+\frac{a}{2M_{2}^{2}}\right) {\mbox{arctanh}}\!\left( \beta
_{N}\right) -\frac{a}{2M_{2}^{2}}\beta _{N}\,\,\ln \left[ \frac{M_{2}}{m}%
\right] \right] ,  \label{fnpp}
\end{equation}
where 
\begin{equation}
\beta _{N}=\left( 1-\frac{4m^{2}M_{2}^{2}}{a^{2}}\right)
^{1/2},\,\,\,\,\,\,\,a=2E\left( E_{2}-\beta \left| \vec{p}_{2}\right|
y_{0}\right) =2l.p_{2},\,  \label{betan}
\end{equation}
and 
\begin{equation}
\Delta _{V}=x_{2}^{+}-x_{2}^{-}=\frac{a\beta _{N}}{\left( l+p_{2}\right) ^{2}%
},\,\,\,\,\,\,\,\,x_{2}^{\pm }=\frac{m^{2}+\frac{1}{2}\left( 1\pm \beta
_{N}\right) a}{\left( l+p_{2}\right) ^{2}}.
\end{equation}
The contribution due to the Coulomb interaction in the neutral HSD is
included in Eq.~(\ref{virn}). The infrared divergence which takes place in
Eqs.~(\ref{pt2}) and (\ref{ficou}) when $\lambda \,\rightarrow 0$ in the
virtual radiative correction is canceled by another one of the same kind
which appears in the bremsstrahlung correction, see Ref.~\cite{93}. $\lambda 
$ is the small mass which is assigned to the photon to cut off the
contribution of very soft photons.

In general, the $\Phi _{s}(E)$ and $\Phi _{s}^{\prime }(E)$ are model
independent functions (do not depend on the form factors) that contain terms
of order $\alpha q/\pi M_{1}.$

\section{BIDIMENSIONAL DISTRIBUTIONS}

The evaluation of the differential decay rate that gives the DP with RC of
process in Eq.~(\ref{uno}) is performed by standard trace calculations,
leaving as the relevant independent variables the energies $E_{2}$ and $E$
of the emitted baryon and the electron, respectively. The phase space
integration is accomplished according to the kinematical limits given in
Eqs.~(\ref{klim1}) and (\ref{klim2}), for the TBR and FBR respectively.

\subsection{Four body region}

In this region, the result up to order $\alpha q/\pi M_{1}\;$for the HSD is
compactly given by products of coefficients $H_{i}^{s\prime }$ that depend
on the form factors and model independent functions $\theta _{i}^{sT}$ as 
\begin{equation}
d\Gamma _{B}^{s}\left( \ A^{s}\rightarrow Be\nu \,\gamma \right) dE=\frac{%
\alpha }{\pi }d\Omega \sum_{i=0}^{17}H_{i}^{s\prime }\,\theta
_{i}^{sT}\,,\,\,\,\,\,\,\,\,d\Omega =\frac{G_{v}^{2}}{2}\,\frac{1}{2\pi ^{3}}%
M_{1}dE_{2}dE,  \label{gen1}
\end{equation}
where the coefficients $H_{i}^{s\prime }$ (see Appendix A) are given by 
\begin{equation}
H_{1}^{c}=H_{1}^{n}=H_{1}^{\prime },\,\,H_{i}^{c}=H_{i}^{\prime
},\,\,H_{i}^{n}=H_{i}^{\prime }+N_{i}^{\prime },\,\,\mbox{for \quad }%
i=0,2,...,17,\,\,H_{17}^{\ \prime }=0.
\end{equation}
\noindent The $H_{i}^{s}\,\,$'s do depend on the form factors through
functions $Q_{i}$, $i=1,..,4$. The explicit form of the $Q_{i}\,$'s (Ref.%
\cite{89}) is also given in Appendix~A. The model independent functions are: 
\begin{equation}
\,\,\theta _{i}^{cT}=\theta _{i}^{nT}=\theta _{i}^{T},\quad \mbox{for \quad }%
i=0,2,...,17.
\end{equation}
The $\theta _{1}^{cT},\theta _{1}^{nT},$ and $\theta _{i}^{T}\,$'s with $%
i=0,...,17$ are shown in the Appendix B.

The specific form of Eq.~(\ref{gen1}) for the cases that correspond to the
semileptonic decay of a charged charged HSD$\,(s=c)\,$ or a neutral neutral
HSD$\,(s=n)$ decaying hyperon, respectively, becomes:

\noindent For the charged HSD process, such as $\Sigma
^{-}(p_{1})\rightarrow n(p_{2})+e^{-}(\ell )+\bar{\nu}_{e}(p_{\nu })+\gamma
\left( k\right) $, 
\begin{equation}
d\Gamma _{B}^{c}\left( \ A^{-}\rightarrow B^{0}e^{-}\bar{\nu}_{e}\,\gamma
\right) \ dE=\frac{\alpha }{\pi }d\Omega \sum_{i=0}^{16}H_{i}^{\prime
}\,\theta _{i}^{T}\,,  \label{gench}
\end{equation}
where the model-independent function $\theta _{1}^{cT}$ is 
\begin{equation}
\theta _{1}^{T}=\left( I_{1}-2\right) \,\ln \left| \frac{y_{0}+1}{y_{0}-1}%
\right| ,\,\,\,I_{1}={\frac{{2}}{{\beta }}\,\mbox{arctanh}}\left( {\beta }%
\right) {,}\,\,\,y_{0}=\frac{\left( E_{\nu }^{0}\right) ^{2}-E^{2}\beta
^{2}-\left| \vec{p}_{2}\right| ^{2}}{2\left| \vec{p}_{2}\right| E\beta },\,\,%
\mbox{and }\,\,E_{\nu }^{0}=M_{1}-E_{2}-E.  \label{tet1c}
\end{equation}
\noindent For the neutral HSD case,{\bf \ }such{\bf \ }as the {\bf %
\thinspace \thinspace \thinspace }$\Lambda (p_{1})\rightarrow
p^{+}(p_{2})+e^{-}(\ell )+\bar{\nu}_{e}(p_{\nu })+\gamma \left( k\right) ,$ 
\begin{equation}
d\Gamma _{B}^{n}\left( \ A^{0}\rightarrow B^{+}e^{-}\bar{\nu}\;\gamma
\right) dE=\frac{\alpha }{\pi }d\Omega \left[ \left( H_{0}^{\prime
}+N_{0}^{\prime }\right) \theta _{0}^{T}+H_{1}^{\prime }\theta
_{1}^{nT}+\sum_{i=2}^{16}\left( H_{i}^{\prime }+N_{i}^{\prime }\right)
\theta _{i}^{T}+N_{17}^{\prime }\theta _{17}^{T}\right] ,  \label{genn}
\end{equation}
where $\theta _{1}^{nT}$ is separated, for further convenience, into two
parts as follows: 
\begin{equation}
\theta _{1}^{nT}=\theta _{1}^{nTD}+\theta _{1}^{nTND},\mbox{ where }%
\,\,\,\,\,\,\,\,\,\,\,\,\,\,\,\,\,\theta _{1}^{nTD}\left( E,E_{2}\right)
=-2\ln \left| \frac{y_{0}+1}{y_{0}-1}\right| .  \label{tet1n}
\end{equation}
The$\;$equations for the $\theta _{1}^{nTND}$ and $\theta _{17}^{T}$ are
also included in the Appendix~B.

The result for both types of unpolarized decays given in Eq.~(\ref{gen1}) is
of high precision, model independent and useful for processes where the
momentum transfer is not small and therefore cannot be neglected. Terms of
order $\alpha q^{2}/\pi M_{1}^{2}\ $and higher are neglected.

\subsection{Three body region}

The result up to order $\alpha q/\pi M_{1}\;$for the hyperon semileptonic
decay (HSD) in this region is given by 
\begin{equation}
d\Gamma _{TBR}^{s}\left( \ A^{s}\rightarrow Be\nu \,\right) dE=\left[
A_{0}^{\prime }+\frac{\alpha }{\pi }\left( \sum_{i=0}^{17}H_{i}^{s\prime
}\,\theta _{i}^{\prime \,\,s}+A_{_{V}}^{s}\Phi _{s}^{\prime }\right) \right]
d\Omega \,,\,\,\,\,\,\,\,d\Omega =\frac{G_{{\rm v}}^{2}}{2}\,\frac{1}{2\pi
^{3}}M_{1}M_{2}\,dE\,dz,\,\,\,\,\,\,z=\frac{E_{2}}{M_{2}}.  \label{gen1tb}
\end{equation}
\noindent The coefficients $H_{i}^{s}\,\,$'s are the same as the ones at the
FBR (Appendix A), but the model independent functions are different, here 
\[
\,\,A_{_{V}}^{c}=B_{1}^{\prime \prime
},\,\,\,\,\,\,\,\,A_{_{V}}^{n}=A_{1N}^{\prime \prime
},\,\,\,\,\,\,\,\,\,\,\,\theta _{1}^{\prime \,\,s}=\Phi _{s}+\theta
_{1}^{s},\,\,\,\,\,\,\,\,\,\theta _{i}^{\prime \,\,c}=\theta _{i}^{\prime
\,\,n}=\theta _{i},\quad \mbox{for \quad }i=0,2,...,17. 
\]

\noindent The $A_{0}^{\prime }$ and $A_{_{V}}^{s}\,\,$'s (given in Appendix
A) also depend on the form factors through functions $Q_{i}$, $i=1,..,4$.
The model independent functions $\theta _{1}^{\prime \,\,s}$ , $\Phi
_{s}^{\prime }$ and the $\theta _{i}^{s}\,$'s ($i=0,2,...,17\,)$ are
exhibited in Appendix~C.

In order to illustrate the result in Eq.~(\ref{gen1tb}), let us consider two
cases that correspond to the semileptonic decay of a charged and a neutral
decaying hyperon:

\noindent For the charged HSD

\begin{equation}
d\Gamma ^{c}(A^{-}\rightarrow B^{0}e^{-}\overline{\nu })dE=\left[
A_{0}^{\prime }+\frac{\alpha }{\pi }\left( H_{0}^{\prime }\theta
_{0}+H_{1}^{\prime }\left( \Phi _{c}+\theta _{1}^{c}\right) +B_{1}^{\prime
\prime }\Phi _{c}^{\prime }+\sum_{i=2}^{16}H_{i}^{\prime }\theta _{i}\right)
\right] d\Omega ,  \label{b}
\end{equation}
\noindent For the neutral HSD, 
\begin{equation}
d\Gamma ^{n}(A^{0}\rightarrow B^{+}e^{-}\overline{\nu _{e}})\,dE=\left[
A_{0}^{\prime }+\frac{\alpha }{\pi }\left[ \left( H_{0}^{\prime
}+N_{0}^{\prime }\right) \theta _{0}+H_{1}^{\prime }\left( \Phi _{n}+\theta
_{1}^{n}\right) +A_{1N}^{\prime \prime }\Phi _{n}^{\prime
}+\sum_{i=2}^{17}\left( H_{i}^{\prime }+N_{i}^{\prime }\right) \theta
_{i}\right] \right] d\Omega .
\end{equation}
The $\theta _{1}^{\prime \,\,s}$ are analyzed in the Appendix C.

\section{ENERGY SPECTRUM}

To obtain the energy spectrum of the beta particle, the integration over $%
E_{2}$ has to be performed in the corresponding kinematical regions, $\left(
E_{2b}<E_{2}<E_{2t}\right) $, according to Eq.~(\ref{klim1}) and Eq.~(\ref
{klim2}) 
\begin{equation}
\Gamma ^{s}(E)=\int_{E_{2b}}^{E_{2t}}\frac{d\Gamma (A^{s}\rightarrow Be\nu
\,)}{dE}.
\end{equation}
The result for the case without any radiative corrections becomes 
\begin{equation}
\Gamma _{o}^{s}(A^{s}\longrightarrow Be\overline{\nu })=\frac{G_{v}^{2}}{%
2\pi ^{3}}\frac{M_{1}}{2}\int_{E_{2}^{\min }}^{E_{2}^{\max }}A_{0}^{\prime
}\,\,dE_{2},\,\,\,\,\,\,\,\,\,\,\,\mbox{where }\,\,\,\,\,\,\,\,\,\,\,\,\,\,%
\sum_{sp\acute{\imath}n}\left| M_{0}\right| ^{2}=\frac{G_{v}^{2}}{2}\frac{%
4M_{1}}{M_{2}mm_{\nu }}A_{0}^{\prime }.
\end{equation}

\subsection{Pure Bremsstrahlung correction in the FBR}

The $d\Gamma _{B}^{s}(A^{s}\rightarrow Be\nu \,\gamma )\,$ in Eq.~(\ref{gen1}
) depends on $y_{0}$ through $\theta _{1}^{sT}$ [see Eq.~(\ref{tet1c}) and
Eq.~(\ref{tet1n})] in the following way: 
\begin{equation}
\theta _{i}^{sT}\propto \ln \left| \frac{y_{0}+1}{y_{0}-1}\right| .
\end{equation}

For collinear events, which take place when \thinspace 
\begin{equation}
E_{2}\rightarrow E_{2}^{\min },\quad \qquad m\leq E\leq E_{c}
\end{equation}
the $y_{0}\rightarrow 1$. This gives rise to a logarithmic divergence. This
divergence is an obstacle to perform a direct numerical integration.
However, this difficulty is controlled by performing an analytical
integration. To proceed, it is convenient to take the following definitions: 
\begin{equation}
\frac{(y_{0}+1)}{(y_{0}-1)}=\frac{a_{0}+b_{0}z+s}{a_{0}+b_{0}z-s}%
,\,\,\,\,a_{0}=\frac{M_{1}\left( E_{m}-E\right) +M_{2}^{2}}{M_{2}E\beta }%
,\quad \quad b_{0}=\frac{E-M_{1}}{E\beta }\,\,\,\,\mbox{and }\,\text{ }s=%
\sqrt{z^{2}-1}  \label{defab}
\end{equation}
and to consider separately the terms where the difficulties arise.

For the{\bf \thinspace \thinspace }charged HSD the separation of terms in
Eq.~(\ref{gench} ) is such that the ones that contain a logarithmic
divergence (at the integration limits) are accumulated in $I_{C}^{TD}$ and
all the other non-divergent ones are accommodated in $I_{C}^{TND},$ then 
\begin{equation}
\Gamma _{B}^{c}(E)=\frac{G_{v}^{2}}{2}\frac{1}{2\pi ^{3}}M_{1}\left[
I_{C}^{TD}\left( E\right) +I_{C}^{TND}\left( E\right) \right] ,  \label{ech}
\end{equation}
where 
\begin{equation}
I_{C}^{TD}\left( E\right) =\frac{\alpha }{\pi }\int_{M_{2}}^{E_{2}^{\min
}}H_{1}^{\prime }\ \left( E,E_{2}\right) \ \theta _{1}^{cT}\left(
E,E_{2}\right) \ dE_{2}=\frac{\alpha }{\pi }M_{2}\sum_{k=0}^{2}\varepsilon
_{k}^{c}M_{2}^{k}\ \,R_{k}^{TB}.
\end{equation}
The explicit $z$ dependence of $H_{1}^{\prime }\ \left( E,E_{2}\right) $ is
given by 
\begin{equation}
H_{1}^{\prime }=\sum_{k=0}^{2}\varepsilon
_{k}M_{2}^{k}z^{k},\,\,\,\,\,\,\,\varepsilon _{k}^{c}=\left( I_{1}-2\right)
\varepsilon _{k},\,\,\mbox{and }\,\text{\thinspace \thinspace \thinspace
\thinspace }\,\,\,R_{k}^{FB}=\int\limits_{z=1}^{z_{t}^{B}}z^{k}\ln \frac{%
\left| a_{0}+b_{0}z+\sqrt{z^{2}-1}\right| }{\left| a_{0}+b_{0}z-\sqrt{z^{2}-1%
}\right| }dz.  \label{defh1}
\end{equation}
The $\varepsilon _{k}$ 's are shown in Appendix D. The second term in Eq.~(%
\ref{ech}) is 
\[
I_{C}^{TND}\left( E\right) =\frac{\alpha }{\pi }\int_{M_{2}}^{E_{2}^{\min
}}\left( H_{0}^{\prime }\theta _{0}^{T}+\sum_{i=2}^{16}H_{i}^{\prime }\theta
_{i}^{T}\right) dE_{2}. 
\]
For the neutral HSD, we follow the same strategy and we obtain 
\begin{equation}
\Gamma _{B}^{n}(E)=\frac{G_{v}^{2}}{2}\frac{1}{2\pi ^{3}}M_{1}\left[
I_{N}^{TD}\left( E\right) +I_{N}^{TND}\left( E\right) \right]
\end{equation}
where 
\begin{equation}
I_{N}^{TD}\left( E\right) =\frac{\alpha }{\pi }\int_{M_{2}}^{E_{2}^{\min
}}H_{1}^{\prime }\theta _{1}^{nTD}\ dE_{2}=\frac{\alpha }{\pi }%
M_{2}\sum_{k=0}^{2}\varepsilon _{k}^{n}\
M_{2}^{k}R_{k}^{TB},\,\,\,\,\,\,\varepsilon _{k}^{n}=-2\varepsilon _{k}
\label{epsfbn}
\end{equation}
and 
\begin{equation}
I_{N}^{TND}\left( E\right) =\frac{\alpha }{\pi }\int_{M_{2}}^{E_{2}^{\min
}}\left( \left( H_{0}^{\prime }+N_{0}^{\prime }\right) \theta
_{0}^{T}+H_{1}^{\prime }\theta _{1}^{nTND}+\sum_{i=2}^{16}\left(
H_{i}^{\prime }+N_{i}^{\prime }\right) \theta _{i}^{T}+N_{17}^{\prime
}\theta _{17}^{T}\right) dE_{2}.
\end{equation}
Notice that, we obtain the same integral $R_{k}^{TB}$in both cases. After
performing a very subtle analysis (illustrated in Refs.~\cite{m00} and\cite
{fga}) we get the following non-divergent analytical result 
\begin{equation}
R_{k}^{TB}=\frac{1}{k+1}\left[ z_{t}^{k+1}-\left( z_{t}^{B}\right)
^{k+1}\right] T_{0}^{B}-\frac{1}{k+1}\left[ \sum_{r=0}^{k}\left(
s_{t}z_{t}^{r}-s_{t}^{B}\left( z_{t}^{B}\right) ^{r}\right)
\int\limits_{z=1}^{z_{t}^{B}}z^{k-r}\frac{dz}{s}\right] ,  \label{RkB}
\end{equation}
where 
\begin{equation}
T_{0}^{B}=\mbox{ln}\left| \frac{M_{1}\left( E_{m}-E\right) }{E\beta M_{2}}%
\right| ,\qquad s=\sqrt{z^{2}-1},  \label{TyE}
\end{equation}
and $E_{m}$ is given in Eq.~(\ref{klim2}). The values of $z$ and $s$ (see
Eq.~(\ref{klim0}) ) at the boundaries are 
\begin{eqnarray}
z_{t} &=&\frac{1}{2}\left[ \frac{M_{1}-E\left( 1-\beta \right) }{M_{2}}+%
\frac{M_{2}}{M_{1}-E\left( 1-\beta \right) }\right]
,\,\,\,\,\,z_{t}^{B}=z_{b}=\frac{1}{2}\left[ \frac{M_{1}-E\left( 1+\beta
\right) }{M_{2}}+\frac{M_{2}}{M_{1}-E\left( 1+\beta \right) }\right] , 
\nonumber \\
s_{t} &=&\frac{1}{2}\left( \frac{M_{1}-E\left( 1-\beta \right) }{M_{2}}-%
\frac{M_{2}}{M_{1}-E\left( 1-\beta \right) }\right) ,\quad s_{t}^{B}=s_{bi}=%
\frac{1}{2}\left[ \frac{M_{1}-E\left( 1+\beta \right) }{M_{2}}-\frac{M_{2}}{%
M_{1}-E\left( 1+\beta \right) }\right] .  \label{syz}
\end{eqnarray}
An analogous integral is discussed in Appendix E.

\subsection{Virtual plus Bremsstrahlung corrections at the three body region}

Following a parallel development for the charged HSD we obtain 
\begin{equation}
\Gamma ^{c}(E)=\frac{G_{v}^{2}}{2}\frac{1}{2\pi ^{3}}M_{1}\left[
I_{D}^{c}\left( E\right) +I_{ND}^{c}\left( E\right) \right] ,
\end{equation}
\noindent where 
\begin{equation}
I_{D}^{c}\left( E\right) =\frac{\alpha }{\pi }M_{2}\left[
\sum\limits_{k=0}^{2}\varepsilon _{k}^{\prime \,\,c}\ M_{2}^{k}R_{k}\right]
,\,\,\,\,\,\,\,\varepsilon _{k}^{\prime \,\,c}=\left( I_{1}-2\right)
\varepsilon _{k}  \label{epch}
\end{equation}
\noindent and 
\begin{equation}
I_{ND}^{c}\left( E\right) =\int\limits_{E_{2}^{\min }}^{E_{2}^{\max }}\left[
A_{0}^{\prime }+\frac{\alpha }{\pi }\left( H_{1}^{\prime }\theta
_{1}^{cnd}+B_{1}^{\prime \prime }\Phi _{c}^{\prime }+H_{0}^{\prime }\theta
_{0}+\sum_{i=2}^{16}H_{i}^{\prime }\theta _{i}\right) \right] dE_{2},
\label{zef}
\end{equation}
\begin{equation}
\,R_{k}(z)=\int\limits_{z_{b}}^{z_{t}}z^{k}\ln \left| a_{0}+b_{0}z+\sqrt{%
z^{2}-1}\right| dz.  \label{IRk}
\end{equation}
The $\varepsilon _{k}^{\,\,}$ 's are given in Appendix D, and the
calculation of the integrated form of $R_{k}(z)$ $\,$ is illustrated in
Appendix E. A similar procedure is performed for the neutral HSD, and the
following results are obtained:

\begin{equation}
\Gamma ^{n}(E)=\frac{G_{v}^{2}}{2}\frac{1}{2\pi ^{3}}M_{1}\left[
I_{D}^{n}\left( E\right) +I_{ND}^{n}\left( E\right) \right] ,
\end{equation}
\begin{equation}
I_{D}^{n}\left( E\right) =\frac{\alpha }{\pi }M_{2}\ \ \left[ \varepsilon
_{0}^{\prime }R_{0}+\sum_{n=1}^{2}\left( \varepsilon _{n}^{\prime }\
+\varepsilon _{n}^{\prime \prime }\right) R_{n}+\varepsilon _{3}^{\prime
\prime }R_{3}\right] ,  \label{intnd}
\end{equation}
\begin{equation}
I_{ND}^{n}\left( E\right) =\int\limits_{E_{2}^{\min }}^{E_{2}^{\max }}\left[
A_{0}^{\prime }+\frac{\alpha }{\pi }\left( H_{1}^{\prime }\theta
_{1}^{nnd}+A_{1N}^{\prime \prime }\Phi _{N}^{\prime }+\left( H_{0}^{\prime
}+N_{0}^{\prime }\right) \theta _{0}+\sum_{i=2}^{16}H_{i}^{\prime }\theta
_{i}+\sum_{i=2}^{17}N_{i}^{\prime }\theta _{i}\right) \right] dE_{2}.
\label{fndn}
\end{equation}
The $\theta _{1}^{nnd}$ and the other model independent functions are given
in the Appendix C.

\noindent Here we have considered the approximation 
\begin{equation}
\frac{{\mbox{arctanh}}\!\left( \beta _{N}\right) }{\beta _{N}}\approx \frac{%
I_{1}}{2}-y_{0}\left| \overline{p}_{2}\right| \frac{1}{M_{1}\beta }\left( 1-%
\frac{m^{2}}{E^{2}}\frac{I_{1}}{2}\right) =\frac{I_{1}}{2}-\frac{M_{2}a_{0}}{%
M_{1}\beta }\left( 1-\frac{m^{2}}{E^{2}}\frac{I_{1}}{2}\right) -\frac{%
b_{0}M_{2}}{M_{1}\beta }\left( 1-\frac{m^{2}}{E^{2}}\frac{I_{1}}{2}\right) z.
\label{Aprox}
\end{equation}
According to Eq.~(\ref{epsap}), the model dependent coefficients in Eq.~(\ref
{intnd}), defined as $\varepsilon _{k}^{\prime \,\,n},$ become 
\begin{eqnarray}
\varepsilon _{0}^{\prime \,\,n} &=&\varepsilon _{0}^{\prime }=\varepsilon
_{0}\left[ \left( \,I_{1}-2\right) -\left( \frac{M_{2}a_{0}}{M_{1}\beta }%
\right) \left( 1-\frac{m^{2}}{2E^{2}}I_{1}\right) \right] ,  \nonumber \\
\varepsilon _{1}^{\prime \,\,n} &=&\varepsilon _{1}^{\prime }+\varepsilon
_{1}^{\prime \prime }=\varepsilon _{1}\left( \,I_{1}-2\right) -\left(
b_{0}\varepsilon _{0}+M_{2}a_{0}\varepsilon _{1}\right) \frac{1}{M_{1}\beta }%
\left( 1-\frac{m^{2}}{2E^{2}}I_{1}\right) ,  \nonumber \\
\varepsilon _{2}^{\prime \,\,n} &=&\varepsilon _{2}^{\prime }+\varepsilon
_{2}^{\prime \prime }=\varepsilon _{2}\left( \,I_{1}-2\right) -\left(
b_{0}\varepsilon _{1}+M_{2}a_{0}\varepsilon _{2}\right) \frac{1}{M_{1}\beta }%
\left( 1-\frac{m^{2}}{2E^{2}}I_{1}\right) ,  \nonumber \\
\varepsilon _{3}^{\prime \,\,n} &=&\varepsilon _{3}^{\prime \prime
}=-\varepsilon _{2}\ \left( \frac{b_{0}}{M_{1}\beta }\right) \left( 1-\frac{%
m^{2}}{2E^{2}}I_{1}\right) .  \label{epnn}
\end{eqnarray}
The analytical result for the $R_{k}$ 's in Eqs.~(\ref{epch}) and~(\ref
{intnd}) is 
\begin{equation}
R_{k}=\frac{\left( z_{t}^{k+1}-z_{b}^{k+1}\right) }{2\left( k+1\right) }%
T_{R}-\frac{1}{2\left( k+1\right) }\sum_{r=0}^{k}\left[ \frac{%
(z_{t}^{r}+z_{b}^{r})\left( z_{t}^{k-r+1}-z_{b}^{k-r+1}\right) }{\left(
k-r+1\right) }+\left( s_{t}z_{t}^{r}-s_{bi}z_{b}^{r}\right)
\int_{z_{b}}^{z_{t}}\frac{z^{k-r}}{s}dz\right]  \label{Rk}
\end{equation}
where 
\begin{equation}
T_{R}=\left[ \text{ln}\left| \frac{4\left( E_{m}-E\right) ^{2}}{%
M_{2}^{2}\left( 1-\frac{2E}{M_{1}}+\frac{m^{2}}{M_{1}^{2}}\right) }\right|
+2\Theta \left( E-E_{c}\right) \text{ln}\left| \frac{M_{1}\left(
E_{m}-E\right) }{M_{2}E\beta }\right| \right] ,  \label{FTR}
\end{equation}
the $\Theta \left( E-E_{c}\right) $ is the Heaviside function, \thinspace
and the $z_{t}$, $z_{t}^{B}$, $s_{t}$ , and $s_{t}^{B}$ are given in Eq.~(%
\ref{syz}) 
\begin{equation}
z_{b}=z_{t}^{B},\,\,\,\,\,\,\,\,\,\,\,\,\,\,\,\,s_{bi}=-s_{bd}=s_{t}^{B}.
\end{equation}

\section{FINAL RESULTS}

\subsection{Four body region}

Gathering and refining previous results, we obtain the whole spectrum of
events in the FBR.

\noindent The FBR formula for the lepton energy spectrum becomes in general, 
\begin{equation}
d\Gamma _{B}^{s}\left( E\right) =\frac{\alpha }{\pi }\frac{G_{v}^{2}}{2}%
\frac{1}{2\pi ^{3}}M_{1}M_{2}\left[ \sum_{k=0}^{2}\varepsilon
_{k}^{s}M_{2}^{k}R_{k}^{TB}+\int\limits_{z=1}^{z_{t}^{B}}\left(
H_{0}^{s\prime }\theta _{0}^{T}+\sum_{i=2}^{17}H_{i}^{s\prime }\theta
_{i}^{T}+H_{1}^{\prime }\theta _{1}^{nTND}\delta _{s}^{n}\right) dz\right] ,
\label{masgen}
\end{equation}
the $\delta _{s}^{n}$ indicates that the last term appears only in the
neutral HSD case $(s=n)$. The $\varepsilon _{k}^{s}$ is given in Eq.~(\ref
{defh1}) for $s=c,$ and in Eq.~(\ref{epsfbn}) for the $s=n$ case.

For $k=0,1,2$ the explicit values for $R_{k}^{TB}$, $\,$\thinspace in Eq.~(%
\ref{RkB}) are 
\begin{eqnarray}
R_{0}^{TB} &=&\Delta _{z}T_{0}^{B}-\Delta _{s}\ln \left|
z_{t}^{B}+s_{t}^{B}\right| ,\,\,\,\,\,\,\,\,\,\,\,\,R_{1}^{TB}=\frac{1}{2}%
\left[ z_{t}^{2}-\left( z_{t}^{B}\right) ^{2}\right] T_{0}^{B}-\frac{1}{2}%
\Delta _{s}\ s_{t}^{B}-\frac{1}{2}\left(
s_{t}z_{t}-s_{t}^{B}z_{t}^{B}\right) \ln \left| z_{t}^{B}+s_{t}^{B}\right| ,
\nonumber \\
R_{2}^{TB} &=&\frac{1}{3}\left[ z_{t}^{3}-\left( z_{t}^{B}\right)
^{3}\right] T_{0}^{B}-\frac{1}{3}\left( s_{t}z_{t}-s_{t}^{B}z_{t}^{B}+\Delta
_{s}\frac{z_{t}^{B}}{2}\right) s_{t}^{B}-\frac{1}{3}\left(
s_{t}z_{t}^{2}-s_{t}^{B}\left( z_{t}^{B}\right) ^{2}+\frac{\Delta _{s}}{2}%
\right) \ln \left| z_{t}^{B}+s_{t}^{B}\right| ,
\end{eqnarray}
where $T_{0}^{B}$ is given in Eqs.~(\ref{TyE}), $s_{t},\,z_{t},$ $%
s_{t}^{B},\,z_{t}^{B},$ are given in Eqs.~(\ref{syz}), and 
\begin{equation}
\Delta _{z}=z_{t}-z_{t}^{B}=\frac{2M_{1}\left( E_{m}-E\right) E\beta }{%
M_{2}M_{1}^{2}\left( 1-\frac{2E}{M_{1}}+\frac{m^{2}}{M_{1}^{2}}\right) }%
,\,\,\,\,\,\,\,\,\Delta _{s}=s_{t}-s_{t}^{B}=\frac{E\beta }{M_{2}}\left( 1+%
\frac{M_{2}^{2}}{M_{1}^{2}\left( 1-\frac{2E}{M_{1}}+\frac{m^{2}}{M_{1}^{2}}%
\right) }\right) .  \label{delts}
\end{equation}

\subsection{Three body region}

The TBR formula for the electron energy spectrum becomes 
\begin{equation}
d\Gamma _{B}^{s}\left( E\right) =\frac{G_{v}^{2}}{2}\frac{1}{2\pi ^{3}}%
M_{1}M_{2}\left\{ \int\limits_{z_{b}}^{z_{t}}A_{0}^{\prime }dz+\frac{\alpha 
}{\pi }\left[ \sum_{k=0}^{r_{s}}\varepsilon _{k}^{\prime
\,s}M_{2}^{k}R_{k}+\int\limits_{z_{b}}^{z_{t}}\left( H_{0}^{s\prime }\theta
_{0}+A_{_{V}}^{s}\Phi _{s}^{\prime }+\sum_{i=2}^{17}H_{i}^{s\prime }\theta
_{i}+H_{1}^{\prime }\theta _{1}^{snd}\right) dz\right] \right\}
\label{gentb}
\end{equation}
where $r_{c}=2,\,\,\,\,r_{n}=3$, and the\thinspace $\varepsilon _{k}^{\prime
\,s}$ 's $\,$are displayed in Eq.~(\ref{epch}) and in Eq.~(\ref{epnn}).

The explicit values for $R_{k}$, in Eq.~(\ref{Rk}) for $k=0,1,2,3$ are 
\begin{equation}
R_{0}=\frac{1}{2}\left[ \Delta _{z}\left( T_{R}-2\right) -\Delta _{s}\ \text{
ln}\left| \frac{\ z_{t}+s_{t}}{z_{b}+\left| s_{b}\right| }\right| \right] ,
\end{equation}
\begin{equation}
R_{1}=\frac{1}{4}\left[ \left( z_{t}^{2}-z_{b}^{2}\right) \left(
T_{R}-2\right) -\Delta _{s}\Delta _{\left| s\right| }-\left(
s_{t}z_{t}-s_{bi}z_{b}\right) \text{ln}\left| \frac{z_{t}+s_{t}}{%
z_{b}+\left| s_{b}\right| }\right| \right] ,\,\,\,\,\,\,\,\,\,\,\,\Delta
_{\left| s\right| }=s_{t}-\left| s_{b}\right|
\end{equation}
\begin{eqnarray}
R_{2} &=&\frac{1}{6}\left\{ \left( z_{t}^{3}-z_{b}^{3}\right) \left(
T_{R}-2\right) -\frac{\Delta _{z}^{3}}{6}-\frac{1}{2}\Delta _{s}\left(
z_{t}\ s_{t}-z_{b}\left| s_{b}\right| \right) \right\}  \nonumber \\
&&-\frac{1}{6}\left\{ \left( s_{t}z_{t}-s_{bi}z_{b}\right) \Delta _{\left|
s\right| }+\left[ s_{t}z_{t}^{2}-s_{bi}z_{b}^{2}+\frac{1}{2}\Delta
_{s}\right] \text{ln}\left| \frac{z_{t}+s_{t}}{z_{b}+\left| s_{b}\right| }%
\right| \right\} ,
\end{eqnarray}
\begin{eqnarray}
R_{3} &=&\frac{1}{8}\left[ \left( z_{t}^{4}-z_{b}^{4}\right) \left(
T_{R}-2\right) -\frac{\Delta _{z}^{3}}{3}\left( z_{t}+z_{b}\right) \right] -%
\frac{1}{8}\left[ \frac{2}{3}\Delta _{s}\Delta _{\left| s\right| }+\frac{1}{3%
}\Delta _{s}\left( s_{t}z_{t}^{2}-\left| s_{b}\right| z_{b}^{2}\right)
+\left( s_{t}z_{t}^{2}-s_{bi}z_{b}^{2}\right) \Delta _{\left| s\right|
}\right]  \nonumber \\
&&-\frac{1}{8}\left[ \frac{1}{2}\left( s_{t}z_{t}-s_{bi}z_{b}\right) \left(
z_{t}\ s_{t}-z_{b}\ \left| s_{b}\right| \right) +\left(
s_{t}z_{t}^{3}-s_{bi}z_{b}^{3}+\frac{1}{2}\left(
s_{t}z_{t}-s_{bi}z_{b}\right) \right) \text{ln}\left| \frac{z_{t}+s_{t}}{%
z_{b}+\left| s_{b}\right| }\right| \right] ,
\end{eqnarray}
where $T_{R}$ is given in Eq.~(\ref{FTR}), and $s_{t},\,z_{t},$ $%
s_{bi},\,z_{b},$ $\Delta _{z}$, and $\Delta _{s}$ are given in Eqs.~(\ref
{syz}), and Eqs.~(\ref{delts}), respectively.

\section{CONCLUSIONS}

The singularities contained in the precise analytical formulas at the level
of the Dalitz plot are isolated and well understood. The logarithmic
divergences disappear and the lepton energy spectrum becomes finite when the
formulas are integrated in an analytical way. The final and general results
in Eq.~(\ref{masgen}) and Eq.~(\ref{gentb}) are precise formulas suitable to
be evaluated numerically, without ambiguity, at any energy in which the
charged lepton is emitted. They contain the bremsstrahlung in the four body
decay region where the three body decay does not take place and also include
events in which the electron is collinear to the produced hadron (at the
edge of the DP).

Other authors Ref.~\cite{Toth2} have published numerical data for the
percent contributions due to the {\it radiative corrections} to the HSD
decay. Only as a test, we have compared booth results. We consider the
numerical values obtained with the formulas in Eq.~(\ref{masgen}) and Eq.~(%
\ref{gentb}) in six Tables. Tables I-III are devoted to the charged HSD
process, considering the case : $\Sigma ^{-}\rightarrow n+e+\bar{\nu}_{e}$ ,
and Tables IV-VI are entailed to the neutral HSD{\bf \ }case{\bf , }the
process we use to illustrate this case is $\Lambda \rightarrow p+e^{-}+\bar{%
\nu}_{e}$. In the Tables I, III, IV and VI we exhibit the data given in Ref.~%
\cite{Toth2} for the relative RC in \%, caused by bremsstrahlung events,
which fall inside and outside the TBR Dalitz plot, and the numerical values
obtained by means of Eq.~(\ref{masgen}) and Eq.~(\ref{gentb}). The results
are in good agreement taking into account the fact that the earlier
numerical results contain unspecified approximations especially at the
Dalitz plot boundary. In addition we illustrate our final results with four
graphs. For the $\left( \Sigma ^{-},n\right) $ case, in Fig.~\ref{fig1}, the
RC at the TBR (which contains the virtual and bremsstrahlung corrections)
and the RC in FBR (which contains only the bremsstrahlung corrections)
contributions are displayed separately. In Fig.~\ref{fig2}, the complete
energy spectrum is plotted with and without the total radiative corrections.
For the $\left( \Lambda ,p\right) $ case, Fig.~\ref{fig3} contains the
radiative contributions including the Coulomb effect and in Fig.~\ref{fig4}
we consider the energy spectrum as in the previous case. The Coulomb effect
is included in the analysis. It is incorporated into the virtual radiative
corrections to the energy spectrum for the neutral HSD in the TBR.

In summary, the analytical results are useful to obtain information derived
from experimental data, about the underlying interactions in the decay
processes, the basic symmetries, and the internal structure of hadrons,
through the derivation of precise values of the form factors involved in the
effective interaction.

The knowledge of the energy spectrum of the electron is fundamental for the
determination of the decay rate in these processes. For the complete
determination of the decay rate it is necessary to add the contributions of
the events in the three body and in the four body regions.

Let us mention that the RC were also computed by other means for photon
bremsstrahlung calculations in semileptonic decays Ref.~\cite{Gluck}. At
last, though the evaluation of the radiative corrections for the HSD is a
complex and an old problem (see list of references in Ref.~\cite{Gluck}),
the results in Eqs.~(\ref{masgen}) and (\ref{gentb}) are new, and are the
culmination of a systematical approach to the decays in the whole region of
the Dalitz plot.

\acknowledgments

We thank Professor P.~Kielanowski for reviewing the manuscript and providing
valuable suggestions. S.R.J.W. also acknowledges partial support by
Comisi\'{o}n de Operaci\'{o}n y Fomento de Actividades Acad\'{e}micas
(Instituto Polit\'{e}cnico Nacional), and by CoNaCyT-M\'{e}xico.

\begin{center}
\appendix
{\bf APPENDIX A}
\end{center}

The coefficients $Q_{i}$'s $\left( i=1,...,5\right) $ do depend on the form
factors. 
\begin{eqnarray*}
Q_{1} &=&F_{1}^{2}\left( \frac{2E_{2}-M_{2}}{M_{1}}\right) +\frac{1}{2}%
F_{2}^{2}\left( \frac{M_{2}+E_{2}}{M_{1}}\right) +F_{1}F_{2}\left( \frac{%
M_{2}+E_{2}}{M_{1}}\right) +F_{1}F_{3}\left( 1+\frac{M_{2}}{M_{1}}\right)
\left( 1-\frac{E_{2}}{M_{1}}\right) \\
&&+F_{2}F_{3}\left( \frac{M_{2}+E_{2}}{M_{1}}\right) \left( 1-\frac{E_{2}}{%
M_{1}}\right) +G_{1}^{2}\left( \frac{2E_{2}+M_{2}}{M_{1}}\right) -\frac{1}{2}%
G_{2}^{2}\left( \frac{M_{2}-E_{2}}{M_{1}}\right) +G_{1}G_{2}\left( \frac{%
M_{2}-E_{2}}{M_{1}}\right) \\
&&+G_{1}G_{3}\left( \frac{M_{2}}{M_{1}}-1\right) \left( 1-\frac{E_{2}}{M_{1}}%
\right) -G_{2}G_{3}\left( \frac{M_{2}-E_{2}}{M_{1}}\right) \left( 1-\frac{%
E_{2}}{M_{1}}\right) +M_{1}^{2}Q_{5}\left[ \left( \frac{M_{1}-E_{2}}{M_{1}}%
\right) ^{2}-\frac{1}{2}\frac{q^{2}}{M_{1}^{2}}\right] ,
\end{eqnarray*}
\begin{eqnarray*}
Q_{2} &=&-\frac{F_{1}^{2}}{M_{1}}-\frac{F_{1}F_{2}}{M_{1}}+\frac{F_{1}F_{3}}{%
M_{1}}\left( 1+\frac{M_{2}}{M_{1}}\right) +\frac{F_{2}F_{3}}{M_{1}}\left( 
\frac{M_{2}+E_{2}}{M_{1}}\right) -\frac{G_{1}^{2}}{M_{1}}+\frac{G_{1}G_{2}}{%
M_{1}}+\frac{G_{1}G_{3}}{M_{1}}\left( \frac{M_{2}}{M_{1}}-1\right) \\
&&\ \ \ -\frac{G_{2}G_{3}}{M_{1}}\left( \frac{M_{2}-E_{2}}{M_{1}}\right) +%
\frac{2F_{1}G_{1}}{M_{1}}+M_{1}Q_{5}\left( \frac{M_{1}-E_{2}}{M_{1}}\right) ,
\end{eqnarray*}
\[
Q_{3}=Q_{1}-2F_{1}^{2}\left( \frac{E_{2}-M_{2}}{M_{1}}\right)
-2G_{1}^{2}\left( \frac{E_{2}+M_{2}}{M_{1}}\right) -M_{1}^{2}Q_{5}\left[
\left( 1-\frac{E_{2}}{M_{1}}\right) ^{2}-\frac{q^{2}}{M_{1}^{2}}\right] , 
\]
\begin{equation}
Q_{4}=Q_{2}-4\frac{F_{1}G_{1}}{M_{1}},\,\,\,\,\,\,\,\,\,\,\,\,\,\,\,\,\,\,\,%
\,\,\,\,\,\,\,\,\,\,\,\,\,Q_{5}=\frac{F_{3}^{2}}{M_{1}^{2}}\left( \frac{%
M_{2}+E_{2}}{M_{1}}\right) -\frac{G_{3}^{2}}{M_{1}^{2}}\left( \frac{%
M_{2}-E_{2}}{M_{1}}\right) -2\frac{F_{1}F_{3}}{M_{1}^{2}}+2\frac{G_{1}G_{3}}{%
M_{1}^{2}},  \eqnum{A1}  \label{Qs}
\end{equation}
with 
\begin{eqnarray}
F_{1} &=&f_{1}^{\prime }+\left( 1+\frac{M_{2}}{M_{1}}\right) f_{2},\mbox{ }%
F_{2}=-2f_{2},\mbox{ }F_{3}=f_{2}+f_{3},  \nonumber \\
G_{1} &=&g_{1}^{\prime }-\left( 1-\frac{M_{2}}{M_{1}}\right) g_{2},\mbox{ }%
G_{2}=-2g_{2},\mbox{ }G_{3}=g_{2}+g_{3},  \eqnum{A2}  \label{FG}
\end{eqnarray}
$f_{1}^{\prime }$, $g_{1}^{\prime }$ are effective form factors, see Ref.~%
\cite{89}.\vspace{0.1in}

For completeness, we explicitly show the coefficients $H_{i}^{\prime }$ $^{%
\mbox{,}}$s: 
\[
A_{0}^{\prime }=Q_{1}^{\prime }EE_{\nu }^{0}-Q_{2}^{\prime }E\left| \vec{p}%
_{2}\right| \left( \left| \vec{p}_{2}\right| +\left| \vec{l}\right|
y_{0}\right) -Q_{3}^{\prime }\left| \vec{l}\right| \left( \left| \vec{p}%
_{2}\right| y_{0}+\left| \vec{l}\right| \right) +Q_{4}^{\prime }E_{\nu
}^{0}\left| \vec{p}_{2}\right| \left| \vec{l}\right| y_{0}-Q_{5}^{\prime
}\left| \vec{p}_{2}\right| ^{2}\left| \vec{l}\right| y_{0}\left( \left| \vec{%
p}_{2}\right| +\left| \vec{l}\right| y_{0}\right) , 
\]
\[
H_{0}^{\prime }=E\beta \left| \vec{p}_{2}\right| \left[ \frac{1}{2}\left(
Q_{3}-Q_{4}E_{\nu }^{0}\right) -\frac{E}{2M_{1}}\left[ \left(
f_{1}+g_{1}\right) ^{2}+4f_{2}g_{1}+2\left( f_{1}f_{3}-g_{1}g_{2}\right)
\right] \right] , 
\]
\[
H_{1}^{\prime }=A_{1N}^{\prime }=E\,\left\{ E_{\nu }^{0}\,Q_{1}-Q_{2}\left| 
\vec{p}_{2}\right| ^{2}-Q_{3}\beta ^{2}E+\beta \left| \vec{p}_{2}\right|
\,y_{0}\left( E_{\nu }^{0}Q_{4}-Q_{3}-EQ_{2}\right) \right\} , 
\]
\[
H_{2}^{\prime }=\frac{m^{2}}{2E}\beta \left| \vec{p}_{2}\right| \left[
-\left( Q_{1}+Q_{3}\right) E_{\nu }^{0}+\left( Q_{2}+Q_{4}\right) E\left| 
\vec{p}_{2}\right| \beta \,y_{0}+\right. \left. Q_{4}\left( E+E_{\nu
}^{0}\right) ^{2}+Q_{2}\left| \vec{p}_{2}\right| ^{2}\right] , 
\]
\[
H_{3}^{\prime }=\beta \left| \vec{p}_{2}\right| \left\{ \frac{E}{4}\left[
\left[ 2\,E_{\nu }^{0}-E\left( 1+\beta ^{2}\right) \right] \,\left(
Q_{1}+Q_{3}\right) +\right. \right. 2\beta \left| \vec{p}_{2}\right|
y_{0}\left( E_{\nu }^{0}Q_{4}-Q_{3}-EQ_{2}\right) -2\left| \vec{p}%
_{2}\right| {^{2}}\,Q_{2} 
\]
\[
\left. +\left( E_{\nu }^{0}+E\right) \,\left[ E\,\left( 1+\beta ^{2}\right)
\,\left( Q_{2}+3Q_{4}\right) -2\,\left( Q_{3}+2EQ_{4}\right) \right] \right] 
\]
\[
+m^{2}\left[ \frac{h^{+}}{e}\left( E_{\nu }^{0}+2E\right) +\left[ f_{1}{^{2}}%
+g_{1}{^{2}}+2\left( f_{1}f_{3}-g_{1}g_{2}\right) \right] \frac{E_{\nu }^{0}%
}{2M_{1}}-\right. \left. \left. g_{1}\left( f_{1}+f_{2}+g_{2}\right) \frac{%
\left( 2E+E_{\nu }^{0}\right) }{M_{1}}\right] \right\} , 
\]
\[
H_{4}^{\prime }=\frac{1}{2}E\beta \left| \vec{p}_{2}\right| \left\{ \frac{1}{%
2}E\left[ Q_{1}-E_{\nu }^{0}Q_{2}-E\left( Q_{2}+Q_{4}\right) +3\left(
Q_{3}-E_{\nu }^{0}Q_{4}\right) \right] \right. +\frac{1}{2}\left[
-E^{2}\beta ^{2}Q_{2}+\left| \vec{p}_{2}\right| ^{2}Q_{4}+E_{\nu }^{0}\left(
2Q_{3}-3E_{\nu }^{0}Q_{4}\right) \right] 
\]
\[
-\frac{EE_{\nu }^{0}}{M_{1}}\left[ \left( f_{1}-g_{1}\right)
^{2}+2f_{1}f_{3}\right] +\frac{1}{M_{1}}\left[ f_{1}^{\ 2}-g_{1}^{\
2}+2\left( g_{1}g_{2}+f_{1}f_{3}\right) \right] \left| \vec{p}_{2}\right|
\beta Ey_{0}+\frac{2E}{M_{1}}g_{1}^{\ }g_{2}\left[ 2E_{\nu }^{0}+E\left(
4-3\beta ^{2}\right) \right] 
\]
\[
+\frac{2E}{M_{1}}g_{1}^{\ }\left[ 2E\left( f_{1}+f_{2}-g_{2}\right) \right.
\left. +f_{2}\left( E_{\nu }^{0}+\beta ^{2}E\right) \right] +\frac{2h^{-}}{%
e_{i}}\beta ^{2}E^{2}-\left. 2E\left[ E_{\nu }^{0}+2E\left( 1-\beta
^{2}\right) \right] \frac{h^{+}}{e}\right\} , 
\]
\[
H_{5}^{\prime }=\frac{1}{4}E^{2}\beta ^{2}\left| \vec{p}_{2}\right| \left\{
Q_{1}-\left( E+E_{\nu }^{0}\right) Q_{2}+3Q_{3}-\left( 3E+7E_{\nu
}^{0}\right) Q_{4}\right. -4\left( 2E+E_{\nu }^{0}\right) \frac{h^{+}}{e}%
-8E_{\nu }^{0}\frac{h^{-}}{e}+\frac{4}{M_{1}}\left[ E_{\nu }^{0}\left(
f_{1}^{\;2}+2f_{1}f_{3}\right) \right. 
\]
\[
\left. \left. +\left( 2E-E_{\nu }^{0}\right) \left( f_{1}+f_{2}\right)
g_{1}+\left( 2E+3E_{\nu }^{0}\right) g_{1}g_{2}\right] \right\}
,\,\,\,\,\,\,\,\,H_{6}^{\prime }=\frac{\left| \vec{p}_{2}\right| \left(
1-\beta ^{2}\right) E\beta }{4}\left[ Q_{1}+Q_{3}-\left( Q_{2}+Q_{4}\right)
\left( E_{\nu }^{0}+E\right) \right] , 
\]
\[
H_{7}^{\prime }=-\frac{E\beta \left| \vec{p}_{2}\right| }{4}\left\{ \frac{%
\left( 2E-E_{\nu }^{0}\right) }{E}\left( Q_{1}+Q_{3}\right) +\frac{\left| 
\vec{p}_{2}\right| ^{2}}{E}\left( Q_{2}+Q_{4}\right) \right. +2\left( \beta
^{2}-1\right) EQ_{4}+\left| \vec{p}_{2}\right| \beta y_{0}\left(
Q_{2}+3Q_{4}-2\frac{h^{+}}{e}\right) 
\]
\[
-2\left( E_{\nu }^{0}+E\beta ^{2}\right) \frac{h^{+}}{e}+\left[ \left( \beta
^{2}-3\right) E-2E_{\nu }^{0}\right] Q_{2}+\frac{E}{M_{1}}\left( 1-\beta
^{2}\right) \left[ f_{1}^{2}+g_{1}^{2}-g_{1}\left(
2f_{1}+3f_{2}+g_{2}\right) \right] 
\]
\[
\left. +g_{1}\left( f_{2}-g_{2}\right) \left( \frac{M_{2}}{E}\right) \left( 
\frac{2E_{2}-M_{1}}{M_{2}}-\frac{M_{2}}{M_{1}}\right) +\frac{2m^{2}}{EM_{1}}%
\left( f_{1}f_{3}-g_{1}g_{2}\right) \right\} , 
\]
\[
H_{8}^{\prime }=\frac{E\beta \left| \vec{p}_{2}\right| }{4}\left\{
Q_{1}+Q_{3}-\left( 2E+E_{\nu }^{0}\right) \,Q_{2}+\left( E_{\nu
}^{0}-E\right) \,Q_{4}+\right. 2E_{\nu }^{0}\left( \frac{2h^{-}-h^{+}}{e}%
\right) +\left[ \left( f_{1}-g_{1}\right) ^{2}+2f_{1}f_{3}\right] \left( 
\frac{E-2E_{\nu }^{0}}{M_{1}}\right) 
\]
\[
\left. +g_{1}\frac{\left[ \left( 2f_{2}+g_{2}\right) \left( E_{\nu
}^{0}-2\,E\right) +g_{2}E_{\nu }^{0}\right] }{M_{1}}\right\}
,\,\,\,\,\,\,\,\,\,\,\,\,H_{9}^{\prime }=\frac{\left| \vec{p}_{2}\right| 
{\it \beta }}{8}\left[ -\left( Q_{1}+Q_{3}\right) +\left( Q_{2}+Q_{4}\right)
\left( E+E_{\nu }^{0}\right) \right] , 
\]
\[
H_{10}^{\prime }=\frac{1}{4}\left( E\beta \right) ^{3}\left| \vec{p}%
_{2}\right| \left\{ -\left( Q_{2}+5Q_{4}\right) -4\left[ \frac{3h^{-}+2h^{+}%
}{e}\right] +\right. \left. \frac{2}{M_{1}}\left[ 3f_{1}\left(
f_{1}+2f_{3}\right) +g_{1}\left( g_{1}-4f_{1}-6f_{2}+8g_{2}\right) \right]
\right\} , 
\]
\[
H_{11}^{\prime }=0,\,\,\,\,\,\,\,\,\,\,\,\,\,\,\,H_{12}^{\prime }=\left(
E\beta \right) ^{2}\left| \vec{p}_{2}\right| ^{2}\left\{ \frac{h^{+}}{e}-%
\frac{Q_{4}}{2}+\right. \left. \frac{1}{2M_{1}}\left[ \left(
f_{1}+g_{1}\right) ^{2}+2g_{1}\left( 3f_{2}-2g_{2}\right)
+2f_{1}f_{3}\right] \right\} , 
\]
\[
H_{13}^{\prime }=\frac{\left( E\beta \right) ^{2}\left| \vec{p}_{2}\right|
^{2}}{2}\left\{ Q_{4}-2\frac{h^{+}}{e}+\right. \left. \frac{1}{M_{1}}\left[
2f_{2}\left( f_{1}-g_{1}\right)
+g_{1}^{2}-f_{1}^{2}-2f_{2}^{2}-2f_{1}f_{3}\right] \right\} , 
\]
\[
H_{14}^{\prime }=\frac{\left( E\beta \right) ^{2}\left| \vec{p}_{2}\right| }{%
4M_{1}}\left[ \left( f_{1}-g_{1}\right) ^{2}-4f_{2}g_{1}+2\left(
f_{1}f_{3}-g_{1}g_{2}\right) -M_{1}Q_{2}\right] , 
\]
\begin{equation}
H_{15}^{\prime }=\frac{E\beta \left| \vec{p}_{2}\right| }{8}\left\{ -\left(
Q_{2}+Q_{4}\right) -4\frac{h^{-}}{e}+\frac{2}{M_{1}}\left[ \left(
f_{1}-g_{1}\right) ^{2}-2f_{2}g_{1}+2f_{1}f_{3}\right] \right\}
,\,\,\,\,\,\,H_{16}^{\prime }=\frac{\left| \vec{p}_{2}\right| }{4M_{1}}\beta
\left[ -M_{1}\frac{h^{+}}{e}+g_{1}\left( g_{2}-f_{2}\right) \right] . 
\eqnum{A3}  \label{HS}
\end{equation}

We have used the definition 
\[
h^{\pm }=-g_{1}^{2}\left( \kappa _{1}+\kappa _{2}\right) \pm
f_{1}g_{1}\left( \kappa _{2}-\kappa _{1}\right) .
\]
The coefficients $N_{i}$'s are given by 
\[
N_{0}^{\prime }=-\left| \vec{p}_{2}\right| \frac{E\beta }{2M_{1}}\left[
2\left( E-E_{\nu }^{0}\right) R^{+}+\left( E+2E_{\nu }^{0}\right)
R^{-}+\left( 1-y_{0}\right) \frac{\left| \vec{p}_{2}\right| }{2\beta }%
R^{-}\right] ,\,\,\,\,N_{2}^{\prime }=N_{6}^{\prime }=N_{9}^{\prime
}=N_{11}^{\prime }=N_{15}^{\prime }=0,
\]
\[
N_{3}^{\prime }=\frac{E\beta m^{2}}{M_{1}}\left| \vec{p}_{2}\right|
R^{+},\,\quad N_{4}^{\prime }=-\frac{E\beta }{2M_{1}}\left| \vec{p}%
_{2}\right| \left[ 2E^{2}R^{+}+E\beta \left( E\beta +4\left| \vec{p}%
_{2}\right| y_{0}\right) R^{-}\right] ,\,\,\,\,\,\,\,\,N_{5}^{\prime }=-%
\frac{\left( E\beta \right) ^{2}}{M_{1}}\left| \vec{p}_{2}\right| \left[
ER^{+}+2E_{\nu }^{0}R^{-}\right] ,
\]
\[
N_{7}^{\prime }=\left| \vec{p}_{2}\right| \frac{\beta }{2M_{1}}\left[
2m^{2}+EE_{\nu }^{0}\left( 1-\beta x_{0}\right) \right] R^{+},\,\quad
N_{8}^{\prime }=-\left| \vec{p}_{2}\right| \frac{E\beta }{2M_{1}}\left(
2E+E_{\nu }^{0}\right) R^{+},
\]
\[
N_{10}^{\prime }=-\frac{3\left| \vec{p}_{2}\right| \left( E\beta \right) ^{3}%
}{2M_{1}}R^{-},\qquad N_{12}^{\prime }=\left| \vec{p}_{2}\right| ^{2}\frac{%
E\beta ^{2}}{M_{1}}\left( 2E-E_{\nu }^{0}\right)
R^{+},\,\,\,\,\,\,\,N_{13}^{\prime }=-\left| \vec{p}_{2}\right| ^{2}\frac{%
\left( E\beta \right) ^{2}}{2M_{1}}\left( 2R^{+}-R^{-}\right) ,
\]
\begin{equation}
N_{14}^{\prime }=-\left| \vec{p}_{2}\right| \frac{\left( E\beta \right) ^{2}%
}{M_{1}}R^{+},\,\,\,\,\,\,\,\,\,\,N_{16}^{\prime }=-\left| \vec{p}%
_{2}\right| \frac{\beta }{4M_{1}}R^{+},\quad N_{17}^{\prime }=\left| \vec{p}%
_{2}\right| \frac{E\beta }{4M_{2}}\left[ 2E_{\nu }^{0}+\left( 1-y_{0}\right)
\left| \vec{p}_{2}\right| \beta \right] R^{-},  \eqnum{A4}  \label{NS}
\end{equation}
and 
\begin{equation}
R^{\pm }=\left| f_{1}\right| ^{2}\pm \left| g_{1}\right|
^{2},\,\,\,\,\,\,B_{1}^{\prime \prime }=Q_{1}EE_{\nu }^{0}-Q_{2}E\left| \vec{%
p}_{2}\right| \left( \left| \vec{p}_{2}\right| +\beta Ey_{0}\right) , 
\eqnum{A5}
\end{equation}
\begin{equation}
A_{1N}^{\prime \prime }=\frac{a}{2M_{1}^{2}\left( 1-2E_{\nu
}^{0}/M_{1}\right) }\left[ E_{\nu }^{0}(Q_{1}E_{2}+Q_{4}\vec{p}_{2}^{2})-(%
\vec{p}_{2}^{2}+\left| \vec{p}_{2}\right| \beta
Ey_{0})(Q_{2}E_{2}+Q_{3})\right] .  \eqnum{A6}
\end{equation}

\begin{center}
\appendix
{\bf APPENDIX B}
\end{center}

The FBR model independent functions $\theta _{1}^{cT}$ and $\,\theta
_{1}^{nT}$ are given in Eqs.~(\ref{tet1c}) and (\ref{tet1n}), respectively, 
\begin{equation}
\theta _{0}^{T}=2\,\left( I_{1}-2\right) ,\,\,\,\,\,\,\,\mbox{and\quad }%
\qquad \theta _{i}^{cT}=\theta _{i}^{nT}=\theta _{i}^{T},\quad 
\mbox{for
\quad }i=0,2,...17.  \eqnum{B1}  \label{tetf}
\end{equation}
\begin{equation}
\theta _{1}^{cT}=\theta _{1}^{T}=\left( I_{1}-2\right) \ln \left| \frac{%
y_{0}+1}{y_{0}-1}\right| ,\,\,\,\,\,\,\,\,\theta _{1}^{nT}=\,\theta
_{1}^{nTD}+\theta _{1}^{nTND}\,,\,\,\,\,\,\theta _{1}^{nTD}=-2\ln \left| 
\frac{y_{0}+1}{y_{0}-1}\right| .  \eqnum{B2}  \label{tet1f}
\end{equation}
The FBR $\theta _{1}^{nTND}$ in Eqs.~(\ref{tet1n}) is 
\[
\theta _{1}^{nTND}=\frac{1}{2}\left( \ln ^{2}{{v_{\max }^{+}-}}\ln ^{2}{{\
v_{\min }^{+}}}\right) -\frac{1}{2}\left( \ln ^{2}{{v_{\min }^{-}-}}\ln ^{2}{%
\ {v_{\max }^{-}}}\right) 
\]
\begin{equation}
+\frac{1}{\beta _{N}}\left[ \left. \ln {v}^{+}\ln \left| \frac{{v}%
^{+}-a\left( 1+\beta _{N}\right) }{{v}^{+}-a\left( 1-\beta _{N}\right) }%
\right| \right| _{{v}^{+}{{=\ v_{\min }^{+}}}}^{{v}^{+}{{=\ v_{\max }^{+}}}%
}+\left. \ln {v}^{-}\ln \left| \frac{{v}^{-}-a\left( 1+\beta _{N}\right) }{{v%
}^{-}-a\left( 1-\beta _{N}\right) }\right| \right| _{{v}^{-}{{=\ v_{\min
}^{-}}}}^{{v}^{-}{{=\ v_{\max }^{-}}}}\right] -\frac{1}{\beta _{N}}\left[
I_{1}^{n}-I_{2}^{n}+I_{3}^{n}-I_{4}^{n}\right] .\,  \eqnum{B3}
\label{tet1nf}
\end{equation}
For $i=2,...17$%
\[
\theta _{2}^{T}=\frac{1}{\beta \left| \vec{p}_{2}\right| }\left[ \frac{%
I_{2}^{-}}{1+\beta a^{-}}-\frac{I_{2}^{+}}{1+\beta a^{+}}+\frac{E^{2}}{m^{2}}%
\left( I_{2}^{+}-I_{2}^{-}+\beta \ln \left| \frac{I_{3}^{-}}{I_{3}^{+}}%
\right| \right) \right] +\frac{2I_{1}}{E\left( 1+\beta a^{-}\right) \left(
1+\beta a^{+}\right) }\quad , 
\]
\[
\theta _{3}^{T}=\frac{I_{1}}{\left| \vec{p}_{2}\right| }\ln \left| {{\frac{%
1+\beta {\it a}^{+}}{1+\beta {\it a}^{-}}}}\right| +\frac{1}{\beta \left| 
\vec{p}_{2}\right| }\left( L\left[ \frac{1-\beta }{1+{\it \beta a}^{-}}%
\right] -L\left[ \frac{1-\beta }{1+{\it \beta a}^{+}}\right] +L\left[ \frac{%
1+\beta }{1+{\it \beta a}^{+}}\right] -L\left[ \frac{1+\beta }{1+{\it \beta a%
}^{-}}\right] \right) , 
\]
\[
\theta _{4}^{T}=\frac{1}{\left| \vec{p}_{2}\right| }\left[
a^{+}I_{2}^{+}-a^{-}I_{2}^{-}+\ln \left| \frac{I_{3}^{-}}{I_{3}^{+}}\right|
\right] ,\,\,\,\,\,\,\,\,\,\,\,\,\,\,\,\,\,\,\,\,\,\,\,\,\,\,\theta _{5}^{T}=%
\frac{1}{2\left| \vec{p}_{2}\right| }\left[ \left( 1-a^{+2}\right)
I_{2}^{+}-\left( 1-a^{-2}\right) I_{2}^{-}+4\frac{\left| \vec{p}_{2}\right| 
}{E\beta }\right] , 
\]
\[
\theta _{6}^{T}=2\frac{\left( y_{0}-a^{-}\right) }{\left( 1+\beta
a^{-}\right) ^{2}}\left( I_{2}^{-}+\beta I_{1}\right) -2\frac{\left(
y_{0}+a^{+}\right) }{\left( 1+\beta a^{+}\right) ^{2}}\left( I_{2}^{+}+\beta
I_{1}\right) +2\left[ 2+\beta \left( \frac{y_{0}-a^{-}}{1+{\it \beta a}^{-}}-%
\frac{y_{0}+a^{+}}{1+{\it \beta a}^{+}}\right) \right] I_{4}, 
\]
\[
\theta _{7}^{T}=2\,\left[ 2I_{1}+\frac{y_{0}-a^{-}}{1+\beta a^{-}}\left(
\beta I_{1}+I_{2}^{-}\right) -\frac{y_{0}+a^{+}}{1+\beta a^{+}}\left( \beta
I_{1}+I_{2}^{+}\right) \right] ,\,\,\,\,\,\,\,\,\theta _{8}^{T}=2\,\left[
4\,+\left( y_{0}-a^{-}\right) I_{2}^{-}-\left( y_{0}+a^{+}\right)
I_{2}^{+}\right] , 
\]
\[
\theta _{9}^{T}=24\,E+2\left[ 6\left( E_{\nu }^{0}-E\right) +\beta \left(
G^{T-}+G^{T+}\right) \right] I_{1}+2\left(
G^{T-}I_{2}^{-}+G^{T+}I_{2}^{+}\right) +2\left| \vec{p}_{2}\right| \left[ 
\frac{\left( y_{0}-a^{-}\right) ^{2}I_{3}^{-}}{1+\beta a^{-}}-\frac{\left(
y_{0}+a^{+}\right) ^{2}I_{3}^{+}}{1+\beta a^{+}}\right] , 
\]
\[
\theta _{10}^{T}=\frac{1}{3\left| \vec{p}_{2}\right| }\left[ 2\left(
a^{-2}-a^{+2}\right) -a^{-3}I_{2}^{-}+a^{+3}I_{2}^{+}+\ln \left| \frac{%
I_{3}^{-}}{I_{3}^{+}}\right| \right] ,\,\,\,\,\,\,\,\theta _{11}^{T}=2\left(
I_{4}-I_{1}\right) \frac{1}{\beta \left| \vec{p}_{2}\right| },\qquad \theta
_{12}^{T}=\frac{1}{\beta \left| \vec{p}_{2}\right| }\theta _{0}^{T},\qquad
\theta _{13}^{T}=0, 
\]
\[
\theta _{14}^{T}=2\left[ \left( 2-a^{-}I_{2}^{-}\right) \left(
y_{0}-a^{-}\right) -\left( 2-a^{+}I_{2}^{+}\right) \left( y_{0}+a^{+}\right)
\right] , 
\]
\[
\theta _{15}^{T}=24E_{\nu }^{0}+4\beta E\left[ a^{-}\left(
y_{0}-a^{-}\right) I_{2}^{-}-a^{+}\left( y_{0}+a^{+}\right) I_{2}^{+}\right]
+2\left| \vec{p}_{2}\right| \left[ \left( y_{0}-a^{-}\right)
^{2}I_{3}^{-}-\left( y_{0}+a^{+}\right) ^{2}I_{3}^{+}\right] , 
\]
\[
\theta _{16}^{T}=24E^{2}\left( I_{1}-2\right) +8\left( E_{\nu }^{0\
2}-2E^{2}\beta ^{2}\right) I_{1}+4E\beta \left| \vec{p}_{2}\right| \left[ 
\frac{\left( y_{0}-a^{-}\right) ^{2}}{1+\beta a^{-}}\left( \beta
I_{1}+I_{2}^{-}\right) -\frac{\left( y_{0}+a^{+}\right) ^{2}}{1+\beta a^{+}}%
\left( \beta I_{1}+I_{2}^{+}\right) \right] , 
\]
\begin{equation}
\theta _{17}^{T}=2I_{1},%
\mbox{ \thinspace \thinspace \thinspace \thinspace
\thinspace }\mbox{and\quad }%
\mbox{ \thinspace \thinspace \thinspace
\thinspace }a^{\pm }=\frac{E_{\nu }^{0}\pm \left| \vec{p}_{2}\right| }{%
E\beta },  \eqnum{B4}  \label{tetfi}
\end{equation}
$I_{1}\,\,$is given in Eq.~(\ref{tet1c}),$\quad I_{2}^{\pm },\quad
I_{3}^{\pm },\quad I_{4},\quad $and $\quad G^{T\pm }\quad $ are and $\beta
_{N}$ is defined in Eq.~(\ref{betan}), 
\begin{equation}
I_{2}^{\pm }=\ln \left| \frac{1+a^{\pm }}{a^{\pm }-1}\right|
,\,\,\,I_{3}^{\pm }=\frac{2}{\left( a^{\pm }\right) ^{2}-1},\,\,\,I_{4}=%
\frac{2}{1-\beta ^{2}},\,\,\,G^{T\pm }=\mp \beta \left[ \frac{2Ea^{\pm
}\left( y_{0}\pm a^{\pm }\right) }{\left( 1+\beta a^{\pm }\right) }+\frac{%
\left| \vec{p}_{2}\right| \left( y_{0}\pm a^{\pm }\right) ^{2}}{\left(
1+\beta a^{\pm }\right) ^{2}}\right] ,  \eqnum{B5}  \label{i1-4}
\end{equation}
\[
{{v_{\max }^{\pm }=2}}\left( E_{2}\pm \left| \vec{p}_{2}\right| \right)
\left( E\pm \left| \vec{\ell}\right| \right) ,\,\,\,\,\,\,\,{{v_{\min }^{\pm
}=2}}\left\{ EE_{2}-\left| \vec{p}_{2}\right| \left| \vec{\ell}\right| \pm
\left| E_{2}\left| \vec{\ell}\right| -\left| \vec{p}_{2}\right| E\right|
\right\} , 
\]
\[
a=M_{1}^{2}-H^{2}-q^{2},\,\,\,H^{2}=\left( p_{1}-\ell \right) ^{2},\quad
q^{2}=\left( p_{1}-p_{2}\right) ^{2}. 
\]
The $I_{i}^{n}$'s in Eq.~(\ref{tet1nf}) are defined in terms of the
Heaviside function $\theta \left( x\right) $. Then for $i=1,...,4$ 
\[
I_{i}^{n}=I_{iA}\;\theta \left( r_{Ai}\right) +I_{iB}\;\theta \left(
r_{Bi}\right) \;\theta \left( r_{Bi}^{\prime }\right) +I_{iC}\;\theta \left(
r_{Ci}\right) , 
\]
where 
\[
I_{1A,\,2A}=L\left( \frac{a\left( 1\pm \beta _{N}\right) }{{{v_{\min }^{+}}}}%
\right) -L\left( \frac{a\left( 1\pm \beta _{N}\right) }{{{v_{\max }^{+}}}}%
\right) +\frac{1}{2}\left( \ln ^{2}{{v_{\max }^{+}-}}\ln ^{2}{{v_{\min }^{+}}%
}\right) , 
\]
\[
I_{1B,\,2B}={\frac{{-{{\pi }^{2}}}}{3}-}L\left( \frac{{{v_{\min }^{+}}}}{%
a\left( 1\pm \beta _{N}\right) }\right) -L\left( \frac{a\left( 1\pm \beta
_{N}\right) }{{{v_{\max }^{+}}}}\right) +\frac{1}{2}\ln ^{2}\frac{a\left(
1\pm \beta _{N}\right) }{{{v_{\min }^{+}}}}+\frac{1}{2}\left( \ln ^{2}{{\
v_{\max }^{+}-}}\ln ^{2}{{v_{\min }^{+}}}\right) , 
\]
\[
I_{1C,\,2C}=L\left( \frac{{{v_{\max }^{+}}}}{a\left( 1\pm \beta _{N}\right) }%
\right) -L\left( \frac{{{v_{\min }^{+}}}}{a\left( 1\pm \beta _{N}\right) }%
\right) +\ln \left| a\left( 1\pm \beta _{N}\right) \right| \ln \left| \frac{{%
\ {v_{\max }^{+}}}}{{{v_{\min }^{+}}}}\right| , 
\]
\[
I_{3A,\,4A}=L\left( \frac{a\left( 1\pm \beta _{N}\right) }{{{v_{\min }^{-}}}}%
\right) -L\left( \frac{a\left( 1\pm \beta _{N}\right) }{{{v_{\max }^{-}}}}%
\right) -\frac{1}{2}\left( \ln ^{2}{{v_{\min }^{-}-}}\ln ^{2}{{v_{\max }^{-}}%
}\right) , 
\]
\[
I_{3B,\,4B}={\frac{{{{\pi }^{2}}}}{3}+}L\left( \frac{{{v_{\max }^{-}}}}{%
a\left( 1\pm \beta _{N}\right) }\right) +L\left( \frac{a\left( 1\pm \beta
_{N}\right) }{{{v_{\min }^{-}}}}\right) -\frac{1}{2}\ln ^{2}\frac{a\left(
1\pm \beta _{N}\right) }{{{v_{\max }^{-}}}}+\frac{1}{2}\left( \ln ^{2}{{\
v_{\max }^{-}-}}\ln ^{2}{{v_{\min }^{-}}}\right) , 
\]
\[
I_{3C,\,4C}=L\left( \frac{{{v_{\max }^{-}}}}{a\left( 1\pm \beta _{N}\right) }%
\right) -L\left( \frac{{{v_{\min }^{-}}}}{a\left( 1\pm \beta _{N}\right) }%
\right) +\ln \left| a\left( 1\pm \beta _{N}\right) \right| \ln \left| \frac{{%
\ v}_{\max }^{-}}{{{v_{\min }^{-}}}}\right| , 
\]
and the arguments of the Heaviside function are given by the expressions 
\begin{eqnarray*}
r_{A1,A2} &=&-r_{B1,B2}={{v_{\min }^{+}-}}a\left( 1\pm \beta _{N}\right)
,\,\,\,{\,\,\,\,}r_{B1,B2}^{\prime }=-r_{C1,C2}={{v_{\max }^{+}-}}a\left(
1\pm \beta _{N}\right) {,\,\,\,} \\
r_{A3,A4} &=&-\,r_{B3,B4}={{v_{\max }^{-}-}}a\left( 1\pm \beta _{N}\right)
,\,\,\,\,{\,\,\,}r_{B3,B4}^{\prime }=-r_{C3,C4}={{v_{\min }^{-}-}}a\left(
1\pm \beta _{N}\right) {.}
\end{eqnarray*}
The left (right) subindex in the left hand side (LHS) in the former
equations corresponds to the upper (lower) sign in the right hand side (RHS)
in the same equations. $L\left( x\right) $ is the Spence function.

\begin{center}
\appendix
{\bf APPENDIX C}
\end{center}

The {\bf TBR} $\theta _{1}^{\prime \,\,s}\,\,\,$and $\theta _{i}^{\,\,s}$ in
Eqs.~(\ref{gen1tb}) are 
\begin{equation}
\theta _{1}^{\prime \,\,s}=\Phi _{s}+\theta _{1}^{s}.  \eqnum{C1}
\label{tetpv}
\end{equation}
For the charged HSD, after $\,$we split the $\theta _{1}^{nd}$ and $\theta
_{1}^{nnd}$ contributions, the model-independent function becomes 
\begin{equation}
\theta _{1}^{\prime \,\,c}=\left( I_{1}-2\right) \left[ \ln \left|
a_{0}+b_{0}z+\sqrt{z^{2}-1}\right| \right] +\theta _{1}^{cnd},  \eqnum{C2}
\label{bvc}
\end{equation}
\begin{eqnarray}
\theta _{1}^{cnd} &=&-\frac{1}{\beta }L\left( \frac{2\beta E}{M_{1}-E+\beta E%
}\right) +\frac{I_{1}}{2}\left( 1+\frac{m^{2}}{E\left( M_{1}-2E\right) }%
\right) -\frac{1}{\beta }\ln \left| 1-\frac{2\beta E}{M_{1}-E+\beta E}%
\right| \left( \ln \left| \frac{M_{1}}{m}\right| -\beta \frac{I_{1}}{2}%
\right)  \nonumber \\
&&+\left( 1-\frac{I_{1}}{2}\right) \ln \left| \frac{M_{1}mE_{\nu
}^{0}(M_{1}-E_{2}-E_{m})}{E^{2}\beta ^{2}M_{2}^{2}}\right| -\frac{1}{2\beta }%
\ln \left| \frac{1-\beta x_{0}}{1-\beta }\right| \ln \left| \frac{1-\beta
x_{0}}{1+\beta }\right| +\frac{3}{2}\ln \left| \frac{M_{1}}{m}\right| -\frac{%
11}{8}  \nonumber \\
&&+\frac{1}{\beta }\left[ L\left( \frac{\beta (1-x_{0})}{1-\beta x_{0}}%
\right) +L\left( \frac{\beta (1+x_{0})}{1+\beta }\right) \right] -\frac{1}{4}%
\beta I_{1}^{2},  \eqnum{C3}  \label{bvc2}
\end{eqnarray}
where 
\begin{equation}
x_{0}=-\frac{(\left| \vec{p}_{2}\right| y_{0}+\beta E)}{E_{\nu }^{0}}%
,\,\,\,\,\,1-\beta x_{0}=\frac{M_{1}(M_{1}-E_{2}-E_{m})}{EE_{\nu }^{0}}. 
\eqnum{C4}  \label{ecx0}
\end{equation}
To obtain the results in Eq.~(\ref{intnd}) and in Eq.~(\ref{fndn}), for the
neutral HSD{\bf ,} the $\theta _{1}^{\prime \,\,n}$ is separated for
convenience into a $\theta _{1}^{nd}$ and $\theta _{1}^{nnd}$. These
model-independent functions are 
\begin{equation}
\theta _{1}^{\prime \,\,n}=\theta _{1}^{nd}+\theta
_{1}^{nnd},\,\,\,\,\,\,\,\,\,\,\,\,\,\theta _{1}^{nd}=\theta _{I0}^{\prime
},\,\,\,\,\,\,\,\theta _{1}^{nnd}=\Phi _{Coulomb}+C_{N}^{\prime }. 
\eqnum{C5}
\end{equation}
After considering the approximation in Eq.~(\ref{Aprox}) 
\begin{equation}
\theta _{I0}^{\prime }=\psi _{D1}+E_{2}\ \psi _{D2},  \eqnum{C6}
\end{equation}
and 
\begin{equation}
H_{1}^{\prime }\ \theta _{I0}^{\prime }\ =\sum_{n=0}^{2}\varepsilon
_{n}M_{2}^{n}z^{n}\left[ \psi _{D1}+E_{2}\ \psi _{D2}\right] =\left[
\sum_{n=0}^{2}\varepsilon _{n}^{\prime }\
M_{2}^{n}\,z^{n}+\sum_{m=1}^{3}\varepsilon _{m}^{\prime \prime }\
M_{2}^{m}z^{m}\right] \ln \left| a_{0}+b_{0}z+\sqrt{z^{2}-1}\right| , 
\eqnum{C7}
\end{equation}
where 
\begin{equation}
\psi _{D1}=\left[ I_{1}-2-\frac{M_{2}a_{0}}{M_{1}\beta }\left( 1-\frac{m^{2}%
}{E^{2}}\frac{I_{1}}{2}\right) \right] \ln \left| a_{0}+b_{0}z+\sqrt{z^{2}-1}%
\right| ,  \eqnum{C8}
\end{equation}
\begin{equation}
\psi _{D2}=-\frac{b_{0}}{M_{1}\beta }\left( 1-\frac{m^{2}}{E^{2}}\frac{I_{1}%
}{2}\right) \ln \left| a_{0}+b_{0}z+\sqrt{z^{2}-1}\right| ,  \eqnum{C9}
\end{equation}
\begin{equation}
\varepsilon _{n}^{\prime }=\varepsilon _{n}\left[ I_{1}-2-\frac{M_{2}a_{0}}{%
M_{1}\beta }\left( 1-\frac{m^{2}}{E^{2}}\frac{I_{1}}{2}\right) \right]
,\,\,\,\,\varepsilon _{m}^{\prime \prime }=-\varepsilon _{m-1}\ \frac{b_{0}}{%
M_{1}\beta }\left( 1-\frac{m^{2}}{E^{2}}\frac{I_{1}}{2}\right) .  \eqnum{C10}
\label{epsap}
\end{equation}
\noindent The $\Phi _{Coulomb}$ is given in Eq.~(\ref{ficou}) and 
\begin{equation}
C_{N}^{\prime }=C_{N}+2\left( \frac{{\mbox{arctanh}}\!\left( \beta
_{N}\right) }{\beta _{N}}-1\right) \ln \left| 2M_{2}E\beta \right| +\frac{2{%
\ \mbox{arctanh}}~\left( \beta _{N}\right) }{\beta _{N}}\ln \left( \frac{%
1+\beta }{\beta }\right) ,  \eqnum{C11}
\end{equation}
\begin{eqnarray}
C_{N}=\frac{1}{\beta _{N}} &&\left\{ -\left( {\mbox{arctanh}}~\beta
_{N}\right) ^{2}+L\left( \frac{2a\,\beta _{N}}{a\left( 1+\beta _{N}\right)
+2m^{2}}\right) +L\left( \frac{2a\,\beta _{N}}{a\left( 1+\beta _{N}\right)
+2M_{2}^{2}}\right) -2L\left( -\frac{\left( 1-\beta _{N}\right) }{2\,\beta
_{N}}\right) \right.  \nonumber \\
&&\ +2L\left( -\frac{2E\left| \vec{p}_{2}\right| \left( 1+\beta \right)
\left( 1+y_{0}\right) }{a\left( 1+\beta _{N}\right) }\right) +2L\left( -%
\frac{a\left( 1-\beta _{N}\right) }{2E\left( 1+\beta \right) \left(
E_{2}+\left| \vec{p}_{2}\right| \right) +\left( \beta _{N}-1\right) a}\right)
\nonumber \\
&&\ -\frac{1}{2}L\left( -\frac{a\left( 1+\beta _{N}\right) }{\omega _{0}}%
\right) +\frac{1}{2}L\left( -\frac{a\left( 1-\beta _{N}\right) }{\omega _{0}}%
\right) +\ln \frac{2mM_{2}}{\omega _{\min }^{2}\omega _{0}}{\mbox{arctanh}}%
~\left( \beta _{N}\right)  \nonumber \\
&&\ -\ln \frac{2E\left( 1+\beta \right) \left( E_{2}+\left| \vec{p}%
_{2}\right| \right) +\left( \beta _{N}-1\right) a}{2a\,\beta _{N}}\ln \frac{%
\left[ 2E\left( 1+\beta \right) \left( E_{2}+\left| \vec{p}_{2}\right|
\right) +\left( \beta _{N}-1\right) a\right] a\,\beta _{N}}{2m^{2}M^{2}} 
\nonumber \\
&&\ \left. -\ln \left( \frac{a\left( 1+\beta _{N}\right) +2M_{2}^{2}}{%
a\left( 1-\beta _{N}\right) +2M_{2}^{2}}\right) \left[ \ln \frac{M_{2}}{m}-{%
\ \mbox{arctanh}}~(\beta _{N})+\frac{1}{2}\ln \left( \frac{a\left( 1+\beta
_{N}\right) +2M_{2}^{2}}{a\left( 1-\beta _{N}\right) +2M_{2}^{2}}\right)
\right] \right\}  \nonumber \\
&&+\left( \frac{1}{\beta _{N}}{\mbox{arctanh}}\!(\beta _{N})\right) \left[
-\ln \left( \frac{m^{2}}{a\beta _{N}}\right) +\frac{2M_{2}^{2}+\left(
1+\beta _{N}^{2}\right) a}{2\left( p_{2}+\ell \right) ^{2}}\right] +\left( 
\frac{1}{2}-\frac{m^{2}}{\left( p_{2}+\ell \right) ^{2}}\right) \ln \left( 
\frac{M_{2}}{m}\right)  \nonumber \\
&&-\frac{11}{8}+\ln \left( H^{2}-M_{2}^{2}\right) \left( q^{2}-m^{2}\right)
+\left( \ln \frac{E\left( 1+\beta \right) \left( E_{2}+\left| \vec{p}%
_{2}\right| \right) }{mM_{2}}\right) ^{2}-\left( {\mbox{arctanh}}~\left(
\beta _{N}\right) \right) ^{2},  \eqnum{C12}
\end{eqnarray}
where 
\begin{eqnarray*}
\omega _{0} &=&\omega _{1}+\sqrt{\omega _{1}^{2}-4m^{2}M_{2}^{2}}%
,\,\,\,\,\,\omega _{1}=\frac{ab_{r}-4m^{2}M_{2}^{2}}{a-b_{r}}%
,\,\,\,\,\,\,b_{r}=\frac{M_{2}^{2}\left( q^{2}-m^{2}\right) }{H^{2}-M_{2}^{2}%
}+\frac{m^{2}\left( H^{2}-M_{2}^{2}\right) }{q^{2}-m^{2}}, \\
q^{2}
&=&M_{1}^{2}+M_{2}^{2}-2E_{2}M_{1},\,\,\,\,\,%
\,H^{2}=M_{1}^{2}+m^{2}-2EM_{1},\,\,\,\,\,\,\,\omega _{\min }=\left( 1+\beta
_{N}\right) \sqrt{\frac{\left( a-b_{r}\right) \left( H^{2}-M_{2}^{2}\right)
\left( q^{2}-m^{2}\right) }{a\beta _{N}^{3}}}.
\end{eqnarray*}
Now,

\begin{equation}
\theta _{0}=\left( 1+y_{0}\right) \left( I_{1}-2\right) ,\,\,\,\,\,\,\,\,%
\mbox{and \quad }\,\,\,\,\,\theta _{i}=\frac{1}{\left| \vec{p}_{2}\right| }%
(T_{i}^{+}+T_{i}^{-})\quad \mbox{for \quad }i=0,2,...,16,\,\,\,\theta
_{17}=\left( 1+y_{0}\right) I_{1}  \eqnum{C13}
\end{equation}
\[
T_{2}^{\pm }=\pm \frac{1\mp a^{\pm }}{(1\pm \beta )(1+\beta a^{\pm })}\ln
\left( \frac{1\mp \beta }{1-\beta x_{0}}\right) \pm \frac{(1\pm x_{0})\ln
(1\pm x_{0})}{(1\pm \beta )(1-\beta x_{0})}\pm \frac{1\pm a^{\pm }}{(1\mp
\beta )(1+\beta a^{\pm })}\ln (1\pm a^{\pm })-\frac{(x_{0}+a^{\pm })\ln (\pm
x_{0}\pm a^{\pm })}{(1+\beta a^{\pm })(1-\beta x_{0})}, 
\]

\begin{eqnarray*}
T_{3}^{\pm } &=&\frac{1}{2\beta }\left[ L\left( \frac{1-\beta }{1-\beta
x_{0} }\right) -L\left( \frac{1-\beta x_{0}}{1+\beta }\right) -L\left( \frac{
1+\beta a^{-}}{1-\beta x_{0}}\right) +L\left( \frac{1+\beta a^{-}}{1+\beta }
\right) \right. \\
&&\left. +L\left( \frac{1-\beta x^{0}}{1+\beta a^{+}}\right) -L\left( \frac{
1-\beta }{1+\beta a^{+}}\right) +\ln \left( \frac{1-\beta x_{0}}{1-\beta }
\right) \ln \left( \frac{1+\beta a^{+}}{1+\beta }\right) \right] ,
\end{eqnarray*}

\[
T_4^{\pm }=(x_0\pm 1)\ln (1\pm x_0)\pm (1\pm a^{\pm })\ln (1\pm a^{\pm
})-(x_0+a^{\pm })\ln (\pm x_0\pm a^{\pm }), 
\]

\[
T_{5}^{\pm }=\frac{1}{2}\left[ (1-x_{0}^{2})\ln (1\pm x_{0})+(x_{0}\mp
1)a^{\pm }+1\right. \left. -(1-a^{\pm 2})\ln (1\pm a^{\pm
})+(x_{0}^{2}-a^{\pm 2})\ln [\pm (x_{0}+a^{\pm })]\right] , 
\]

\begin{eqnarray*}
T_{6}^{\mp } &=&\left( -\beta E+\left| \vec{p}_{2}\right| \pm \frac{\beta
E_{\nu }^{0}(x_{0}+a^{\mp })}{1+\beta a^{\mp }}\right) I_{4}\pm \frac{\beta
E_{\nu }^{0}(x_{0}+a^{\mp })}{(1+\beta a^{\mp })^{2}}I_{1}+\left( E_{\nu
}^{0}-\frac{\beta E_{\nu }^{0}(x_{0}+a^{\mp })}{(1+\beta a^{\mp })}\right)
J_{4}-\frac{\beta E_{\nu }^{0}(x_{0}+a^{\mp })}{(1+\beta a^{\mp })^{2}}J_{1}
\\
&&\pm \frac{E_{\nu }^{0}(x_{0}+a^{\mp })}{(1+\beta a^{\mp })^{2}}I_{2}^{\mp
}-\frac{E_{\nu }^{0}(x_{0}+a^{\mp })}{(1+\beta a^{\mp })^{2}}J_{2}^{\mp },
\end{eqnarray*}

\[
T_{7}^{\pm }=\left( \left| \vec{p}_{2}\right| -\beta E\mp \frac{\beta E_{\nu
}^{0}(x_{0}+a^{\pm })}{(1+\beta a^{\pm })}\right) I_{1}\mp \frac{E_{\nu
}^{0}(x_{0}+a^{\pm })}{(1+\beta a^{\pm })}I_{2}^{\pm }+\left( E_{\nu }^{0}- 
\frac{\beta E_{\nu }^{0}(x_{0}+a^{\pm })}{(1+\beta a^{\pm })}\right) J_{1}- 
\frac{E_{\nu }^{0}(x_{0}+a^{\pm })}{(1+\beta a^{\pm })}J_{2}^{\pm }, 
\]

\[
T_{8}^{\pm }=-2(\beta E-\left| \vec{p}_{2}\right| +E_{\nu }^{0}x_{0})\mp
E_{\nu }^{0}(x_{0}+a^{\pm })I_{2}^{\pm }-E_{\nu }^{0}(x_{0}+a^{\pm
})J_{2}^{\pm }, 
\]

\begin{eqnarray*}
\frac{T_{9}^{\pm }}{4\left( \beta E\right) ^{2}} &=&-\frac{3E}{2\left( \beta
E\right) ^{2}}(\beta E-\left| \vec{p}_{2}\right| +E_{\nu }^{0}x_{0})+\left( 
\frac{3(\beta E-\left| \vec{p}_{2}\right| )}{4\beta ^{2}E}+\frac{3E_{\nu
}^{0}\left| \vec{p}_{2}\right| }{4\left( \beta E\right) ^{2}}+\beta G^{\pm
}\right) I_{1}\mp \frac{E_{\nu }^{02}(x_{0}+a^{\pm })^{2}}{4\left( \beta
E\right) ^{2}(1+\beta a^{\pm })}I_{3}^{\pm } \\
&&-\frac{E_{\nu }^{02}(x_{0}+a^{\pm })^{2}}{4\left( \beta E\right)
^{2}(1+\beta a^{\pm })}J_{3}^{\pm }+\left( -\frac{3E_{\nu }^{0}}{4\beta ^{2}E%
}+\frac{3E_{\nu }^{0}(E_{\nu }^{0}+\beta Ex_{0})}{4\left( \beta E\right) ^{2}%
}\pm \beta G^{\pm }\right) J_{1}\pm G^{\pm }J_{2}^{\pm }+G^{\pm }I_{2}^{\pm
},
\end{eqnarray*}

\begin{eqnarray*}
T_{10}^{\mp } &=&\frac{1}{3}(x_{0}^{3}\mp 1)\ln (1\mp x_{0})+\frac{1}{3}%
(a^{\mp ^{3}}\mp 1)\ln (1\mp a^{\mp })-\frac{1}{3}(x_{0}^{3}+a^{\mp
^{3}})\ln (\mp (x_{0}+a^{\mp })) \\
&&+\frac{1}{6}(1-x_{0}^{2})(a^{\mp }+1)-\frac{1}{3}(x_{0}\pm 1)(1-a^{\mp
^{2}}),
\end{eqnarray*}

\[
T_{11}^{\pm }=\frac{1}{2\left| \vec{p}_{2}\right| \beta }\left( E_{\nu
}^{0}\left[ (1-\beta x_{0})J_{4}-J_{1}\right] -\left( \beta E_{\nu
}^{0}+\beta E-\left| \vec{p}_{2}\right| \right) I_{4}+(\beta E-\left| \vec{p}%
_{2}\right| )I_{1}\right) , 
\]

\[
T_{12}^{+}=T_{12}^{-}=\frac{1}{2\left| \vec{p}_{2}\right| \beta }\left(
E_{\nu }^{0}(1-\beta x_{0})J_{1}+2E_{\nu }^{0}x_{0}+2(\beta E-\left| \vec{p}%
_{2}\right| )-(\beta E_{\nu }^{0}+\beta E-\left| \vec{p}_{2}\right|
)I_{1}\right) , 
\]

\[
T_{13}^{\pm }=-\frac{1}{2\left| \vec{p}_{2}\right| }E_{\nu
}^{0}(1-x_{0}^{2}),\,\,\,\,\,\,\,\,\,\,\,\,\,\,\,T_{14}^{\pm }=E_{\nu
}^{0}\left( 1+x_{0}^{2}+2a^{\pm }(x_{0}\mp 1)\pm a^{\pm }(x_{0}+a^{\pm
})(I_{2}^{\pm }\pm J_{2}^{\pm })\right) , 
\]

\[
T_{15}^{\pm }=3E_{\nu }^{0}(2\left| \vec{p}_{2}\right| (1+y_{0})+\beta
E(1-x_{0}^{2}))-E_{\nu }^{0^{2}}(x_{0}+a^{\pm })^{2}(J_{3}^{\pm }\pm
I_{3}^{\pm })-2\beta EE_{\nu }^{0}(x_{0}+a^{\pm })a^{\pm }(J_{2}^{\pm }\pm
I_{2}^{\pm }), 
\]

\begin{eqnarray*}
T_{16}^{\pm } &=&4\beta ^{2}E^{2}\left\{ \frac{3}{2\beta ^{2}}\left( 2(\beta
E-\left| \vec{p}_{2}\right| +E_{\nu }^{0}x_{0})+\beta E_{\nu
}^{0}(1-x_{0}^{2})\right) \right. +\left( -\frac{3(\beta E-\left| \vec{p}%
_{2}\right| +\beta E_{\nu }^{0})}{2\beta ^{2}}-\left| \vec{p}_{2}\right|
(1+y_{0})+\frac{\left| \vec{p}_{2}\right| E_{\nu }^{0^{2}}}{2\beta ^{2}E^{2}}%
\right) I_{1} \\
&&-\frac{E_{\nu }^{0^{2}}(x_{0}+a^{\pm })^{2}}{2\beta E(1+\beta a^{\pm })}%
\left( \beta J_{1}+J_{2}^{\pm }\pm \beta I_{1}\pm I_{2}^{\pm }\right) \left.
+\left( \frac{3E_{\nu }^{0}(1-\beta x_{0})}{2\beta ^{2}}+\frac{E_{\nu
}^{0^{2}}(E_{\nu }^{0}+\beta Ex_{0})}{2\beta ^{2}E^{2}}\right) J_{1}\right\}
.
\end{eqnarray*}

\noindent The $x_{0}$, $a^{\pm }$, $I_{1}$, $I_{2}^{\pm }$, $I_{3}^{\pm }$, $%
I_{4},$ are given in Eq.~(\ref{ecx0}), Eq.~(\ref{tetfi}), Eq.~(\ref{tet1c})
and Eqs.~(\ref{i1-4}). The $J_{1}$, $J_{2}^{\pm }$, $J_{3}^{\pm }$, $J_{4}$
and $G^{\pm }$ are

\[
J_{1}=-\frac{1}{\beta }\left[ \ln \left( \frac{1+\beta }{1-\beta x_{0}}
\right) +\ln \left( \frac{1-\beta }{1-\beta x_{0}}\right) \right]
,\;\,\,\,\,\,\,J_{2}^{\pm }=\ln \left| \frac{a^{\pm }-1}{a^{\pm }+x_{0}}
\right| +\ln \left| \frac{a^{\pm }+1}{a^{\pm }+x_{0}}\right|
,\;\,\,\,\,\,\,J_{3}^{\pm }=-2\left( \frac{a^{\pm }}{a^{\pm ^{2}}-1}-\frac{1 
}{a^{\pm }+x_{0}}\right) , 
\]

\[
J_{4}=\frac{2}{\beta }\left( \frac{1}{1-\beta ^{2}}-\frac{1}{1-\beta x_{0}}%
\right) ,\,\,\,G^{\pm }=\mp \frac{\beta E_{\nu }^{02}(x_{0}+a^{\pm })^{2}}{%
4\beta ^{2}E^{2}(1+\beta a^{\pm })^{2}}\mp \frac{a^{\pm }(a^{\pm 2}-1)}{%
4(1+\beta a^{\pm })}. 
\]
Finally, simplifying Eq.~(\ref{fcpp}) and Eq.~(\ref{fnpp}) 
\[
\Phi _{c}^{\prime }=\frac{m^{2}}{E\left( M_{1}-2E\right) }\left[ \left( 1-%
\frac{M_{1}}{E}\right) \frac{I_{1}}{2}+\ln \left( \frac{M_{1}}{m}\right)
\right] ,\,\,\,\,\,\Phi _{n}^{\prime }=\frac{-2m^{2}}{a}\left[ \left( 1+%
\frac{2M_{2}^{2}}{a}\right) \frac{1}{\beta _{N}}{\mbox{arctanh}\,}\!(\beta
_{N})+\ln \left( \frac{M_{2}}{m}\right) \right] . 
\]

\begin{center}
\appendix
{\bf APPENDIX D}
\end{center}

\begin{equation}
H_{1}^{\prime }=\sum_{k=0}^{2}\varepsilon _{k}E_{2}^{k},  \eqnum{D1}
\end{equation}
\begin{eqnarray*}
\frac{\varepsilon _{0}}{M_{1}^{2}} &=&\left(
F_{1}^{2}-G_{1}^{2}+F_{1}F_{2}+G_{1}G_{2}+\frac{F_{2}^{2}-G_{2}^{2}}{2}%
\right) \left[ M^{-}+4\frac{E}{M_{1}}\left( 1-\frac{E}{M_{1}}\right) \right] 
\frac{M_{2}}{2M_{1}}-2\left( F_{1}^{2}-G_{1}^{2}\right) \left( 1-\frac{E}{%
M_{1}}\right) \frac{M_{2}E}{M_{1}^{2}} \\
&&+\left[ \left( F_{1}+G_{1}\right) ^{2}+F_{1}F_{2}-G_{1}G_{2}\right] \left[
-\frac{1}{2}M^{+}+\frac{E}{M_{1}}\left( 2\left( 1-\frac{E}{M_{1}}\right) +%
\frac{m^{2}}{M_{1}^{2}}\right) \right] +2F_{1}G_{1}\frac{E}{M_{1}}\left[
2\left( \frac{M_{2}^{2}}{M_{1}^{2}}+\frac{E}{M_{1}}\right) -M^{+}\right] \\
&&+\left( F_{1}F_{3}+F_{2}F_{3}+G_{1}G_{3}-G_{2}G_{3}\right) \left( 1-\frac{E%
}{M_{1}}\right) \frac{m^{2}M_{2}}{M_{1}^{3}}+\left(
F_{1}F_{3}-G_{1}G_{3}\right) \left( 1-\frac{E}{M_{1}}\right) \frac{m^{2}}{%
M_{1}^{2}}, \\
\frac{\varepsilon _{1}}{M_{1}} &=&\left( F_{1}F_{2}+G_{1}G_{2}+\frac{%
F_{2}^{2}-G_{2}^{2}}{2}\right) \left( 1-\frac{2E}{M_{1}}\right) \frac{M_{2}}{%
M_{1}}+\left( F_{1}^{2}-G_{1}^{2}\right) \frac{M_{2}}{M_{1}}+\left(
F_{2}F_{3}+G_{2}G_{3}\right) \left( 1-\frac{2E}{M_{1}}\right) \frac{m^{2}}{%
2M_{1}^{2}} \\
&&+\left( {F_{2}^{2}+G_{2}^{2}}\right) \left( 1-\frac{E}{M_{1}}\right) \frac{%
E}{M_{1}}+\left[ \left( F_{1}+G_{1}\right) ^{2}+F_{1}F_{2}-G_{1}G_{2}\right] 
\frac{1}{2}\left( M^{+}+2\left( 1-2\frac{E}{M_{1}}\right) \right) \\
&&+\left( F_{1}F_{2}-G_{1}G_{2}+\frac{F_{2}^{2}+G_{2}^{2}}{2}\right) \frac{1%
}{2}M^{-}-\left( F_{1}F_{3}+F_{2}F_{3}+G_{1}G_{3}-G_{2}G_{3}\right) \frac{%
m^{2}M_{2}}{M_{1}^{3}} \\
&&+\left[ F_{2}F_{3}+G_{2}G_{3}+2\left( G_{1}G_{3}-F_{1}F_{3}\right) \right] 
\frac{m^{2}}{2M_{1}^{2}}, \\
\varepsilon _{2} &=&\left( \frac{F_{2}^{2}+G_{2}^{2}}{2}\right) \left( 1-%
\frac{2E}{M_{1}}\right) -\left( F_{1}+G_{1}\right) ^{2}-\left(
F_{2}F_{3}+G_{2}G_{3}\right) \frac{m^{2}}{M_{1}^{2}}.
\end{eqnarray*}
Terms of $O(q^{2}/M_{1}^{2})$ are neglected and 
\[
M^{-}=\frac{m^{2}-M_{2}^{2}-M_{1}^{2}}{M_{1}^{2}}\quad \text{y\quad }M^{+}=%
\frac{m^{2}+M_{2}^{2}+M_{1}^{2}}{M_{1}^{2}}. 
\]

\begin{center}
\appendix
{\bf APPENDIX E}
\end{center}

In this appendix we describe the integration procedure we follow to obtain
the result in Eq. (\ref{Rk}), Ref.~\cite{tjmc}.

The first step is an integration by parts 
\[
\int z^{n}\text{ln}\left| a_{0}+b_{0}z+\sqrt{z^{2}-1}\right| dz=\frac{z^{n+1}%
}{n+1}\text{ln}\left| a_{0}+b_{0}z+\sqrt{z^{2}-1}\right| -\frac{1}{n+1}\int
D^{(n)}dz, 
\]
where 
\[
D^{(n)}=\frac{z^{n+1}}{s}\left( \frac{b_{0}s+z}{a_{0}+b_{0}z+s}\right)
,\,\,\,\,\,\,\,\,\,\,\,\,\,\,\,\,s=\sqrt{z^{2}-1}. 
\]
After multiplying the numerator and denominator of $D^{(n)}$ by $\left(
a_{0}+b_{0}z-s\right) $ it becomes 
\[
D^{(n)}=z^{n+1}\left( \frac{az}{az^{2}+bz+c}+\frac{a_{0}z}{\left(
az^{2}+bz+c\right) s}+\frac{b}{2\left( az^{2}+bz+c\right) }+\frac{b_{0}}{%
\left( az^{2}+bz+c\right) s}\right) , 
\]
\begin{equation}
a=b_{0}^{2}-1,\quad b=2a_{0}b_{0},\quad c=a_{0}^{2}+1.\,\,\,\,\,\,\,\,%
\mbox{Due to \quad }\,\,\,\,az^{2}+bz+c=a(z-z_{t})(z-z_{b}),  \eqnum{E1}
\label{fden}
\end{equation}
it is reduced to a suitable expression which can be integrated very easily: 
\[
D^{(n)}=-\frac{z^{n+1}}{\Delta _{z}}\left( \left( z_{t}+\frac{b}{2a}\right) 
\frac{1}{z_{t}-z}+\left( z_{b}+\frac{b}{2a}\right) \frac{1}{z-z_{b}}+\left( 
\frac{a_{0}}{a}z_{t}+\frac{b_{0}}{a}\right) \frac{1}{s\left( z_{t}-z\right) }%
+\left( \frac{a_{0}}{a}z_{b}+\frac{b_{0}}{a}\right) \frac{1}{s\left(
z-z_{b}\right) }\right) , 
\]

where $z_{t}$ , $z_{b}$ and $\Delta _{z}$ are defined in Eqs.(\ref{syz}) and
(\ref{delts}). $D^{(n)}$ is simplified to 
\[
D^{(n)}=-\frac{z^{n+1}}{2}\left[ \frac{1}{z_{t}-z}-\frac{1}{z-z_{b}}+\frac{1%
}{s}\left( \frac{s_{t}}{(z_{t}-z)}+\frac{s_{bi}}{(z-z_{b})}\right) \right] . 
\]
The particular results 
\[
D^{(0)}=-\frac{1}{2}\left[ -2+\frac{z_{t}}{z_{t}-z}-\frac{z_{b}}{z-z_{b}}+%
\frac{1}{s}\left( -\Delta _{s}+\frac{s_{t}z_{t}}{z_{t}-z}+\frac{s_{bi}z_{b}}{%
z-z_{b}}\right) \right] , 
\]

\[
D^{(1)}=-\frac{1}{2}\left[ -2z-(z_{t}+z_{b})+\frac{z_{t}^{2}}{z_{t}-z}-\frac{%
z_{b}^{2}}{z-z_{b}}+\frac{1}{s}\left( -\Delta _{s}z+(s_{bi}z_{b}-s_{t}z_{t})+%
\frac{s_{t}z_{t}^{2}}{z_{t}-z}+\frac{s_{bi}z_{b}^{2}}{z-z_{b}}\right)
\right] , 
\]
\begin{eqnarray*}
D^{(2)} &=&-\frac{1}{2}\left\{ -2z^{2}-z(z_{t}+z_{b})-(z_{t}^{2}+z_{b}^{2})+%
\frac{z_{t}^{3}}{z_{t}-z}-\frac{z_{b}^{3}}{z-z_{b}}\right. \\
&&\ \ \left. +\frac{1}{s}\left[ -\Delta
_{s}z^{2}+z(s_{bi}z_{b}-s_{t}z_{t})+\left(
s_{bi}z_{b}^{2}-s_{t}z_{t}^{2}\right) +\frac{s_{t}z_{t}^{3}}{z_{t}-z}+\frac{%
s_{bi}z_{b}^{3}}{z-z_{b}}\right] \right\} ,
\end{eqnarray*}
inspire the following relation:

\begin{eqnarray*}
D^{(n)} &=&-\frac{1}{2}\left[ \sum_{r=0}^{n}\left\{
-(z_{t}^{r}+z_{b}^{r})z^{n-r}-\frac{1}{s}\left(
s_{t}z_{t}^{r}-s_{bi}z_{b}^{r}\right) z^{n-r}\right\} \right. \\
&&\ \left. +\frac{z_{t}^{n+1}}{z_{t}-z}-\frac{z_{b}^{n+1}}{z-z_{b}}+\frac{1}{%
s}\left( \frac{s_{t}z_{t}^{n+1}}{z_{t}-z}+\frac{s_{bi}z_{b}^{n+1}}{z-z_{b}}%
\right) \right] .
\end{eqnarray*}
Therefore

\[
\int z^{n}\text{ln}\left| a_{0}+b_{0}z+s\right| \,dz=T_{D}-\frac{1}{2\left(
n+1\right) }\sum_{r=0}^{n}\left\{ \frac{(z_{t}^{r}+z_{b}^{r})z^{n-r+1}}{%
\left( n-r+1\right) }+\left( s_{t}z_{t}^{r}-s_{bi}z_{b}^{r}\right) \int 
\frac{z^{n-r}}{s}dz\right\} , 
\]
where 
\begin{eqnarray*}
T_{D} &=&\frac{z^{n+1}}{n+1}\text{ln}\left| a_{0}+b_{0}z+\sqrt{z^{2}-1}%
\right| \ -\frac{z_{t}^{n+1}}{2\left( n+1\right) }\left[ \text{ln}\left|
\;z_{t}-z\right| \pm \text{ln}\left| \frac{z_{t}z-1\mp s_{t}s}{z_{t}-z}%
\right| \right] \\
&&\ -\frac{z_{b}^{n+1}}{2\left( n+1\right) }\left[ \text{ln}\left|
\;z-z_{b}\right| \mp \frac{s_{bi}}{\sqrt{z_{b}^{2}-1}}\text{ln}\left| \frac{%
1-z_{b}z\pm s\sqrt{z_{b}^{2}-1}}{z-z_{b}}\right| \right] .
\end{eqnarray*}
To take into account the integration limits we consider that 
\[
\sqrt{z_{b}^{2}-1}=\left\{ 
\begin{tabular}{ll}
$s_{bi}$ & $\,\,\,\,\,\,\,\,\,m<E<E_{c},$ \\ 
$s_{bd}=-s_{bi},$ & $\,\,\,\,\,\,\,\,\,\,E_{c}<E<E_{m}.$%
\end{tabular}
\right. 
\]
and obtain

\[
a_{0}+b_{0}z_{b}+\sqrt{z_{b}^{2}-1}=\left\{ 
\begin{tabular}{ll}
$2s_{bi},$ & $\,\,\,\,\,\,\,\,\,\,\,m<E<E_{c},$ \\ 
$0,$ & $\,\,\,\,\,\,\,\,\,\,E_{c}<E<E_{m}.$%
\end{tabular}
\right. 
\]
\[
a_{0}+b_{0}z_{t}+\sqrt{z_{t}^{2}-1}=\left\{ 
\begin{tabular}{ll}
$0,$ & $\,\,\,\,\,\,\,\,\,\,\,m<E<E_{c},$ \\ 
$0,$ & $\,\,\,\,\,\,\,\,\,\,E_{c}<E<E_{m}.$%
\end{tabular}
.\right. 
\]
As a consequence of these results for $m<E<E_{c}$ and $E_{c}<E<E_{m}$ we
consider two regions.

We define 
\[
T_{D}=\left\{ 
\begin{tabular}{ll}
$T_{D}^{<},$ & $\,\,\,\,\,\,\,\,\,\,\,\,\,\,\,m<E<E_{c},$ \\ 
$T_{D}^{>},$ & $\,\,\,\,\,\,\,\,\,\,\,\,\,E_{c}<E<E_{m}.$%
\end{tabular}
\right. 
\]
At the region $m<E<E_{c}$ taking the limit $z\rightarrow z_{t}$

\begin{eqnarray*}
T_{D}^{<}(z_{t}) &=&\frac{z_{t}}{2\left( n+1\right) }\lim\limits_{z%
\rightarrow z_{t}}\left\{ \text{ln}\left| \frac{\left( a_{0}+b_{0}z+\sqrt{%
z^{2}-1}\right) ^{2}}{z_{t}z-1-s_{t}s_{{}}}\right| \right\} \ \ -\frac{%
z_{b}^{n+1}}{2\left( n+1\right) }\left[ \text{ln}\left|
1-z_{b}z_{t}-s_{bi}s_{t}\right| \right] \\
\ &=&\frac{z_{t}^{n+1}}{2\left( n+1\right) }\text{ln}\left| 2\left(
b_{0}s_{t}+z_{t}\right) ^{2}\right| -\frac{z_{b}^{n+1}}{2\left( n+1\right) }%
\left[ \text{ln}\left| 1-z_{b}z_{t}-s_{bi}s_{t}\right| \right] ;
\end{eqnarray*}
for the limit $z\rightarrow z_{b}$%
\begin{eqnarray*}
T_{D}^{<}(z_{b}) &=&\frac{z_{b}^{n+1}}{2\left( n+1\right) }\text{ln}\left|
\left( a_{0}+b_{0}z_{b}+s_{bi}\right) ^{2}\right| \ \ -\frac{z_{t}^{n+1}}{%
2\left( n+1\right) }\text{ln}\left| z_{t}z_{b}-1-s_{t}s_{bi}\right| -\frac{%
z_{b}^{n+1}}{2\left( n+1\right) }\left[ \text{ln}\left|
1-z_{b}^{2}-s_{bi}^{2}\right| \right] \\
\ &=&\frac{z_{b}^{n+1}}{2\left( n+1\right) }\text{ln}\left| \frac{\left(
2s_{bi}\right) ^{2}}{-2s_{bi}^{2}}\right| -\frac{z_{it}^{n+1}}{2\left(
n+1\right) }\text{ln}\left| z_{t}z_{b}-1-s_{t}s_{bi}\right| .
\end{eqnarray*}
The following remains: 
\[
T_{D}^{<}(z_{t})-T_{D}^{<}(z_{b})=\frac{z_{t}^{n+1}}{2\left( n+1\right) }%
\text{ln}\left| 2\left( b_{0}s_{t}+z_{t}\right) ^{2}\left(
z_{t}z_{b}-1-s_{t}s_{bi}\right) \right| -\frac{z_{b}^{n+1}}{2\left(
n+1\right) }\left[ \text{ln}\left| -2\left( 1-z_{b}z_{t}-s_{bi}s_{t}\right)
\right| \right] . 
\]
Similarly for the second region in which $E_{c}<E<E_{m}$%
\begin{eqnarray*}
T_{D}^{>}(z_{t}) &=&\frac{z_{t}^{n+1}}{2\left( n+1\right) }%
\lim\limits_{z\rightarrow z_{t}}\left\{ \text{ln}\left| \frac{\left(
a_{0}+b_{0}z+\sqrt{z^{2}-1}\right) ^{2}}{z_{t}z-1-s_{t}s}\right| \right\} -%
\frac{z_{b}^{n+1}}{2\left( n+1\right) }\left[ \text{ln}\left|
1-z_{b}z_{t}+s_{bd}s_{t}\right| \right] \\
\ &=&\frac{z_{t}^{n+1}}{2\left( n+1\right) }\text{ln}\left| 2\left(
b_{0}s_{t}+z_{t}\right) ^{2}\right| -\frac{z_{b}^{n+1}}{2\left( n+1\right) }%
\left[ \text{ln}\left| 1-z_{b}z_{t}-s_{bi}s_{t}\right| \right] ,
\end{eqnarray*}
and 
\begin{eqnarray*}
T_{D}^{>}(z_{b}) &=&\frac{z_{b}^{n+1}}{2\left( n+1\right) }%
\lim\limits_{z\rightarrow z_{b}}\left\{ \text{ln}\left| \frac{\left(
a_{0}+b_{0}z+\sqrt{z^{2}-1}\right) ^{2}}{1-z_{b}z+s_{bd}s}\right| \right\} \
\ \ \ -\frac{z_{t}^{n+1}}{2\left( n+1\right) }\left[ \text{ln}\left|
z_{t}z_{b}-1-s_{t}s_{bd}\right| \right] \\
\ &=&\frac{z_{b}^{n+1}}{2\left( n+1\right) }\text{ln}\left| -2\left(
b_{0}s_{bd}+z_{b}\right) ^{2}\right| -\frac{z_{t}^{n+1}}{2\left( n+1\right) }%
\left[ \text{ln}\left| z_{t}z_{b}-1-s_{t}s_{bd}\right| \right] .
\end{eqnarray*}
Then

\[
T_{D}^{>}(z_{t})-T_{D}^{>}(z_{b})=\frac{z_{t}^{n+1}}{2\left( n+1\right) }%
\text{ln}\left| 2\left( b_{0}s_{t}+z_{t}\right) ^{2}\left(
z_{t}z_{b}-1-s_{t}s_{bd}\right) \right| -\frac{z_{b}^{n+1}}{2\left(
n+1\right) }\text{ln}\left| \left( 1-z_{b}z_{t}-s_{bi}s_{t}\right) \left(
-2\right) \left( b_{0}s_{bd}+z_{b}\right) ^{2}\right| .\ \ 
\]

Summarizing

\[
T_{D}^{<}(z_{t})-T_{D}^{<}(z_{b})=\frac{\left(
z_{t}^{n+1}-z_{b}^{n+1}\right) }{2\left( n+1\right) }\text{ln}\left| \frac{%
4\left( E_{m}-E\right) ^{2}}{M_{2}^{2}\left( 1-\frac{2E}{M_{1}}+\frac{m^{2}}{%
M_{1}^{2}}\right) }\right| , 
\]
\[
T_{D}^{>}(z_{t})-T_{D}^{>}(z_{b})=\frac{\left(
z_{t}^{n+1}-z_{b}^{n+1}\right) }{2\left( n+1\right) }\left[ \text{ln}\left| 
\frac{4\left( E_{m}-E\right) ^{2}}{M_{2}^{2}\left( 1-\frac{2E}{M_{1}}+\frac{%
m^{2}}{M_{1}^{2}}\right) }\right| +\text{ln}\left| \frac{M_{1}^{2}\left(
E_{m}-E\right) ^{2}}{M_{2}^{2}\overrightarrow{l}^{2}}\right| \right] . 
\]
In general

\[
T_{D}(z_{t})-T_{D}(z_{b})=\frac{\left( z_{t}^{n+1}-z_{b}^{n+1}\right) }{%
2\left( n+1\right) }T_{R} 
\]
where 
\[
T_{R}=\text{ln}\left| \frac{4\left( E_{m}-E\right) ^{2}}{M_{2}^{2}\left( 1-%
\frac{2E}{M_{1}}+\frac{m^{2}}{M_{1}^{2}}\right) }\right| +2\Theta \left(
E-E_{c}\right) \text{ln}\left| \frac{M_{1}\left( E_{m}-E\right) }{%
M_{2}\left| \overrightarrow{l}\right| }\right| 
\]
and $\Theta \left( E-E_{c}\right) $ is the Heaviside function.

Finally, 
\[
R_{n}(z)=\frac{\left( z_{t}^{n+1}-z_{b}^{n+1}\right) }{2\left( n+1\right) }%
T_{R}-\frac{1}{2\left( n+1\right) }\sum_{r=0}^{n}\left[ \frac{%
(z_{t}^{r}+z_{b}^{r})\left( z_{t}^{n-r+1}-z_{b}^{n-r+1}\right) }{\left(
n-r+1\right) }+\left( s_{t}z_{t}^{r}-s_{bi}z_{b}^{r}\right)
\int_{z_{b}}^{z_{t}}\frac{z^{n-r}}{s}dz.\right] 
\]
\begin{figure}[tbp]
\caption{Radiative corrections for the CHSD in the three and four body
regions of the Dalitz plot.}
\label{fig1}
\end{figure}

\begin{figure}[tbp]
\caption{Electron energy spectrum for the CHSD with all RC contributions and
without any.}
\label{fig2}
\end{figure}

\begin{figure}[tbp]
\caption{ Radiative corrections for the NHSD in the three and four body
regions of the Dalitz plot.}
\label{fig3}
\end{figure}

\begin{figure}[tbp]
\caption{ Electron energy spectrum for the NHSD with all RC contributions
and without any.}
\label{fig4}
\end{figure}

\begin{table}[tbp]
\caption{Radiative corrections in the four body region of the Dalitz plot.}%
\begin{tabular}{cddddd}
%\hline\hline
\multicolumn{6}{c}{
$\Sigma^{-}
(p_{1})\rightarrow
n(p_{2})+e^{-}(\ell )+\bar{\nu}_{e}(p_{\nu})+
\gamma \left( k\right)$ } \\
\hline
$x=E/E_{m}$ & 0.1 & 0.2 & 0.3 & 0.4 & 0.5 \\
 \% in Ref.~\cite{Toth2} & 7.8 & 1.5 & 0.5 & 0.1 & 0.02 \\
\% from Eq.~(\ref{masgen}) & 7.776 & 1.533 & 0.458 & 0.135 & 0.019
%\hline
\end{tabular}
\end{table}
\begin{table}[tbp]
\caption{Radiative corrections in the three body region of the Dalitz plot.}%
\begin{tabular}{cddddddddd}
%\hline\hline
\multicolumn{6}{c}{
$\Sigma^{-}
(p_{1})\rightarrow
n(p_{2})+e^{-}(\ell )+\bar{\nu}_{e}(p_{\nu})$ } \\
\hline
$x=E/E_{m}$ & 0.1 & 0.2 & 0.3 & 0.4 & 0.5 & 0.6 & 0.7 & 0.8 & 0.9\\
\% from Eq.~(\ref{gentb}) & 10.442 & 5.729 & 3.329 & 1.614 & 0.167
& -1.220 & -2.647 & -4.291 & -6.597
%\hline
\end{tabular}
\end{table}
\begin{table}[tbp]
\caption{Total radiative corrections.}%
\begin{tabular}{cddddddddd}
%\hline\hline
\multicolumn{6}{c}{
$\Sigma^{-}
(p_{1})\rightarrow
n(p_{2})+e^{-}(\ell )+\bar{\nu}_{e}(p_{\nu})+
\gamma \left( k\right)$ } \\
\hline
$x=E/E_{m}$ & 0.1 & 0.2 & 0.3 & 0.4 & 0.5 & 0.6 & 0.7 & 0.8 & 0.9\\
 \% in Ref.~\cite{Toth2} & 18.3 & 7.1 & 3.6 & 1.4 &- 0.2
 & -1.7 & -3.2 & -5.0 & -7.5 \\
\% from (Eq.~(\ref{masgen})+Eq.~(\ref{gentb})) & 18.218 & 7.262 & 3.787
& 1.749 & 0.186 & -1.220 & -2.647 & -4.291 & -6.597
%\hline
\end{tabular}
\end{table}
\begin{table}[tbp]
\caption{Radiative corrections in the four body region of the Dalitz plot.}%
\begin{tabular}{cddddd}
\multicolumn{6}{c}{
$\Lambda (p_{1})\rightarrow p^{+}(p_{2})+e^{-}(\ell )+\bar{\nu}_{e}(p_{\nu
})+\gamma \left( k\right) $} \\ \hline
$x=E/E_{m}$ & 0.1 & 0.2 & 0.3 & 0.4 & 0.5 \\
\% in Ref.~\cite{Toth2} & 9.5 & 2.3 & 0.8 & 0.25 & 0.02 \\
\% from Eq.~(\ref{masgen}) & 9.277 & 2.200 & 0.765 & 0.242 & 0.024
%\hline
\end{tabular}
\end{table}
\begin{table}[tbp]
\caption{Radiative corrections in the three body region of the Dalitz plot.}%
\begin{tabular}{cddddddddd}
%\hline\hline
\multicolumn{6}{c}{
$\Lambda (p_{1})\rightarrow p^{+}(p_{2})+e^{-}(\ell )+\bar{\nu}_{e}(p_{\nu
}) $} \\ \hline
$x=E/E_{m}$ & 0.1 & 0.2 & 0.3 & 0.4 & 0.5 & 0.6 & 0.7 & 0.8 & 0.9\\
\% from Eq.~(\ref{gentb}) & 9.058 & 5.213 & 3.119 & 1.531
& 0.091 & -1.396 & -2.933 &-4.692 &-7.146
%\hline
\end{tabular}
\end{table}
\begin{table}[tbp]
\caption{Total radiative corrections.}%
\begin{tabular}{cddddddddd}
%\hline\hline
\multicolumn{6}{c}{
$\Lambda (p_{1})\rightarrow p^{+}(p_{2})+e^{-}(\ell )+\bar{\nu}_{e}(p_{\nu
})+\gamma \left( k\right) $} \\ \hline
$x=E/E_{m}$ & 0.1 & 0.2 & 0.3 & 0.4 & 0.5 & 0.6 & 0.7 & 0.8 & 0.9\\
\% in Ref.~\cite{Toth2} &18.2  & 7.1 & 3.5 & 1.4 & -0.3
& -1.8 & -3.4 & -5.2 & -7.7\\
\% from (Eq.~(\ref{masgen})+Eq.~(\ref{gentb})) & 18.335 & 7.413
& 3.884 & 1.773 & 0.115 & -1.396 & -2.933 & -4.692 & -7.146
\end{tabular}
\end{table}

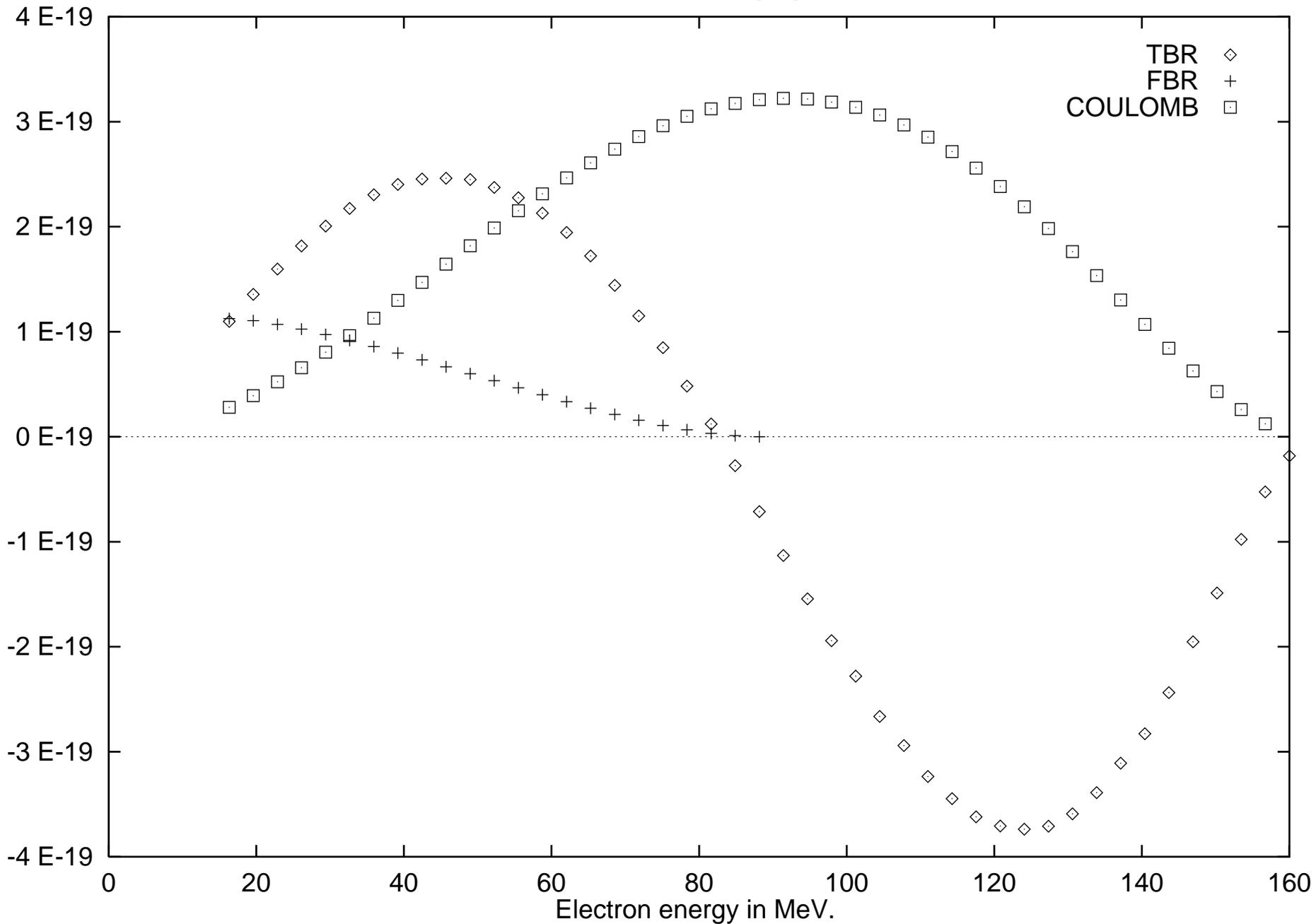
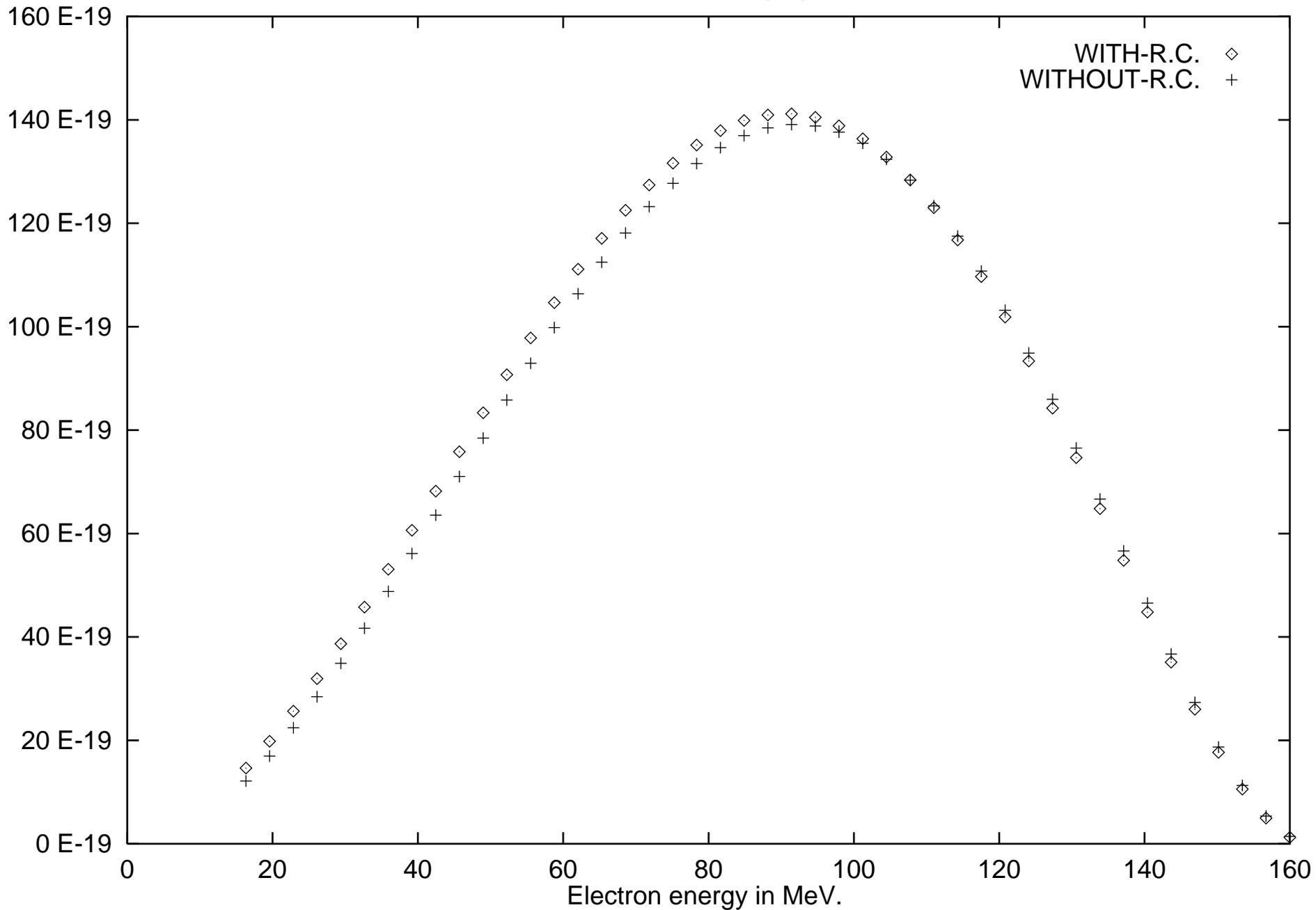
\end{document}